\documentclass[aps,prl,twocolumn,nofootinbib,superscriptaddress]{revtex4}
\usepackage{graphicx,amssymb,amsmath,color,axodraw,dcolumn,subfigure}

\def\CM{{\cal M}}

\newcommand{\ifm}[1]{\relax\ifmmode #1\else $#1$5\fi}
 
\newcommand{\beq}{\begin{equation}}
\newcommand{\eeq}{\end{equation}}
\newcommand{\beqn}{\begin{eqnarray}}
\newcommand{\eeqn}{\end{eqnarray}}
\newcommand{\bi}{\begin{itemize}}
\newcommand{\ei}{\end{itemize}}
\newcommand{\bd}{\begin{description}}
\newcommand{\ed}{\end{description}}
\newcommand{\bHuge}{\begin{Huge}}
\newcommand{\bhuge}{\begin{huge}}
\newcommand{\bLARGE}{\begin{LARGE}}
\newcommand{\bLarge}{\begin{Large}}
\newcommand{\blarge}{\begin{large}}
\newcommand{\eHuge}{\end{Huge}}
\newcommand{\ehuge}{\end{huge}}
\newcommand{\eLARGE}{\end{LARGE}}
\newcommand{\eLarge}{\end{Large}}
\newcommand{\elarge}{\end{large}}

\def \gtsim    {\relax\ifmmode{\mathrel{\mathpalette\oversim >}}
                  \else{$\mathrel{\mathpalette\oversim >}$}\fi}
\def \ltsim    {\relax\ifmmode{\mathrel{\mathpalette\oversim <}}
                  \else{$\mathrel{\mathpalette\oversim <}$}\fi}
\def\oversim#1#2{\lower4pt\vbox{\baselineskip0pt \lineskip1.5pt
            \ialign{$\mathsurround=0pt#1\hfil##\hfil$\crcr#2\crcr\sim\crcr}}}
\def \dk       {\relax\ifmmode{\rightarrow}\else{$\rightarrow$}4\fi}
\def \to       {\relax\ifmmode{\rightarrow}\else{$\rightarrow$}4\fi}
\def \Dk    {\relax\ifmmode{\Rightarrow}\else{$\Rightarrow$}\fi}
\newcounter{minutes}

\newcommand{\met}{\mbox{${E\!\!\!\!/_{\rm T}}$}}

\def \sp       {\relax\ifmmode{\;}\else{$\;$}\fi}	
\def\Journal#1#2#3#4{{#1} {\bf #2}, #3 (#4)}
\def \PRL      {Phys. Rev. Lett.~}
\def \PR       {Phys. Rev.}
\def \PRD      {Phys. Rev. D}

\def \PLB      {Phys. Lett. B}
\def \ZPC      {Z. Phys. C}	
\def \NPB      {Nucl. Phys. B}
\def \PR       {Phys. Rep.~}

\def \NIM      {Nucl. Instrum. Methods}
\def \NIMA     {Nucl. Instrum. Methods Phys. Res. Sect. A}
\def \RPP	{Rep. Prog. Phys.}
\def \PRP	{Prog. Theor. Phys}
\def \MPL	{{Mod. Phys. Lett.} A}
\def \EPJC	{{Eur. Phys. J.} C}
\def \CPC	{Comput. Phys. Commun.}
\def \etal     {$et \; al. \;$}
\begin{document}

\title{LHC discovery potential of the lightest NMSSM Higgs in the $h_1 \to a_1 a_1 \to 4\mu$ channel}
\date{\today}

\author{Alexander Belyaev}
\affiliation{
School of Physics \& Astronomy, University of Southampton,\\
Highfield, Southampton SO17 1BJ, UK}
\affiliation{
Particle Physics Department, Rutherford Appleton Laboratory, \\
Chilton, Didcot, Oxon OX11 0QX, UK}
\author{Jim Pivarski}
\affiliation{Department of Physics and Astronomy, Texas A\&M University, \\
College Station, TX 77843, USA}
\author{Alexei Safonov}
\affiliation{Department of Physics and Astronomy, Texas A\&M University, \\
College Station, TX 77843, USA}
\author{Sergey Senkin}
\affiliation{Department of Physics and Astronomy, Texas A\&M University, \\
College Station, TX 77843, USA}
\author{Aysen Tatarinov}
\affiliation{Department of Physics and Astronomy, Texas A\&M University, \\
College Station, TX 77843, USA}

\begin{abstract}
We explore the potential of the Large Hadron Collider to observe the
$h_1\to a_1a_1\to 4\mu$ signal from the lightest scalar Higgs boson
($h_1$) decaying into the two lightest pseudoscalar Higgs bosons
($a_1$), followed by their decays into four muons in the Next-to-Minimal
Supersymmetric Standard Model (NMSSM). The signature under study
applies to the region of the NMSSM parameter space in which $m_{a_1} <
2 m_\tau$, which has not been studied previously. In such a scenario,
the suggested strategy of searching for a four-muon signal with the
appropriate background suppression would provide a powerful method to
discover the lightest CP-even and CP-odd NMSSM Higgs bosons $h_1$ and
$a_1$.
\end{abstract}

\maketitle

\section{Introduction}

The Next-to-Minimal Supersymmetric Standard Model
(NMSSM)~\cite{Nilles:1982dy,Frere:1983ag,Ellis:1988er,Drees:1988fc,Ellwanger:1993hn,Ellwanger:1993xa,Elliott:1993bs,Pandita:1993tg,Ellwanger:1995ru,King:1995vk,Franke:1995tc,Ellwanger:1996gw,Miller:2003ay}
extends the particle content of the Minimal Supersymmetric Standard
Model (MSSM) by one singlet superfield.  The NMSSM has several
attractive features beyond the MSSM.  First, the NMSSM elegantly
solves the so-called $\mu$ problem~\cite{mu-problem}: the scale of the
$\mu$ parameter is dynamically generated at the electroweak or SUSY
scale when the singlet Higgs acquires a Vacuum Expectation Value
(VEV).  Second, the fine-tuning and little hierarchy problems of the
NMSSM are greatly diminished compared to the
MSSM~\cite{Dermisek:2005ar}.  In the NMSSM, the upper mass limit on
the lightest CP-even Higgs boson is higher than in the MSSM, making it
less constrained experimentally.  Another attractive feature of the
NMSSM is that the lightest CP-even Higgs can have a significant
branching fraction for the new $h_1 \to a_1 a_1$ decay ($h_1$ and
$a_1$ are the lightest CP-even and CP-odd Higgs bosons, respectively).
This weakens the LEP-II constraints on the allowed Higgs parameter
space, as this new decay channel reduces the branching fractions of
$h_1$ into the modes used in direct Higgs searches.  In addition,
there are interesting implications in the cosmological Dark Matter
sector of the model due to the appearance of the fifth neutralino, the
``singlino.''  It has been shown~\cite{nmssm-dm} that the NMSSM is
consistent with the experimentally measured relic density, and the
data provide important constraints on the allowed NMSSM parameters.

The rich phenomenology offered by the NMSSM, stemming from the
extension of the scalar sector, has been the focus of numerous
studies~\cite{nmssm-ph1,nmssm-ph2,nmssm-ph2b,nmssm-ph3,nmssm-ph4,nmssm-ph5,nmssm-ph6,nmssm-ph6a,nmssm-ph7}.
In Ref.~\cite{nmssm-ph2}, the first attempt to establish a ``no-lose''
theorem for NMSSM was presented.  This theorem states that the Large
Hadron Collider (LHC) has the potential to discover at least one NMSSM
Higgs boson in the conventional mode, assuming that Higgs-to-Higgs
decay modes are not important.  However, the fact that Higgs-to-Higgs
decay modes can be important has been shown in analyses devoted to
re-establishing the `no-lose'
theorem~\cite{nmssm-ph2b,nmssm-ph3,nmssm-ph4,nmssm-ph5,nmssm-ph6,nmssm-ph6a,nmssm-ph7}
in the case where the $h_1\to a_1 a_1$ branching fractions are
significant and $a_1$ is light.  NMSSM scenarios with $m_{a_1}$
between the $2\tau$ and $2b$-quark thresholds ($2m_\tau < m_{a_1} <
2m_b$) have previously been considered, focusing on the
$4\tau$~\cite{nmssm-ph7} channels in Higgs-strahlung and Vector Boson
Fusion, establishing the NMSSM No-Lose Theorem at the
LHC~\cite{nmssm-ph7} for this $a_1$ mass region.  Future analysis of
the $4\tau$ channel is likely to be technically challenging and can
only be performed with large datasets (typical integrated luminosity
of 10--100 fb$^{-1}$).  A more recent study~\cite{2mu2tau-pheno} has
focused on the same region using the $h_1 \to a_1 a_1 \to 2\mu 2\tau$
process, which has the benefit that the invariant mass of two muons
forms a much narrower peak, improving the sensitivity of such an
analysis in spite of the large reduction in signal yield due to small
$\mathcal{B}(a_1 \to \mu \mu)$.  Our findings indicate a substantially higher
QCD multi-jet background contamination as compared to
Ref.~\cite{2mu2tau-pheno}, which may have a substantial effect of the
sensitivity of such a search.

In this paper, we explore the region in which the $a_1$ mass is below
the $2\tau$ threshold ($m_{a_1} < 2m_\tau$).  For this case, which has
not been studied previously, we explore the potential of the $h_1 \to
a_1 a_1 \to 4\mu$ signature at the LHC.  Unlike searches for the
$4\tau$ signature, the invariant mass of each muon pair provides a
direct estimate of $m_{a_1}$ and the 4$\mu$ invariant mass provides
$m_{h_1}$.  Use of these kinematic constraints leads to essentially
zero background and therefore allows one to rely on direct $gg$ and
$b\bar{b}$ fusion for Higgs production instead of the subdominant
vector boson fusion or associated production processes chosen in the
$4\tau$ search~\cite{nmssm-ph2} to suppress QCD backgrounds.

\begin{figure*}[thbp]
\includegraphics[width=0.49\linewidth]{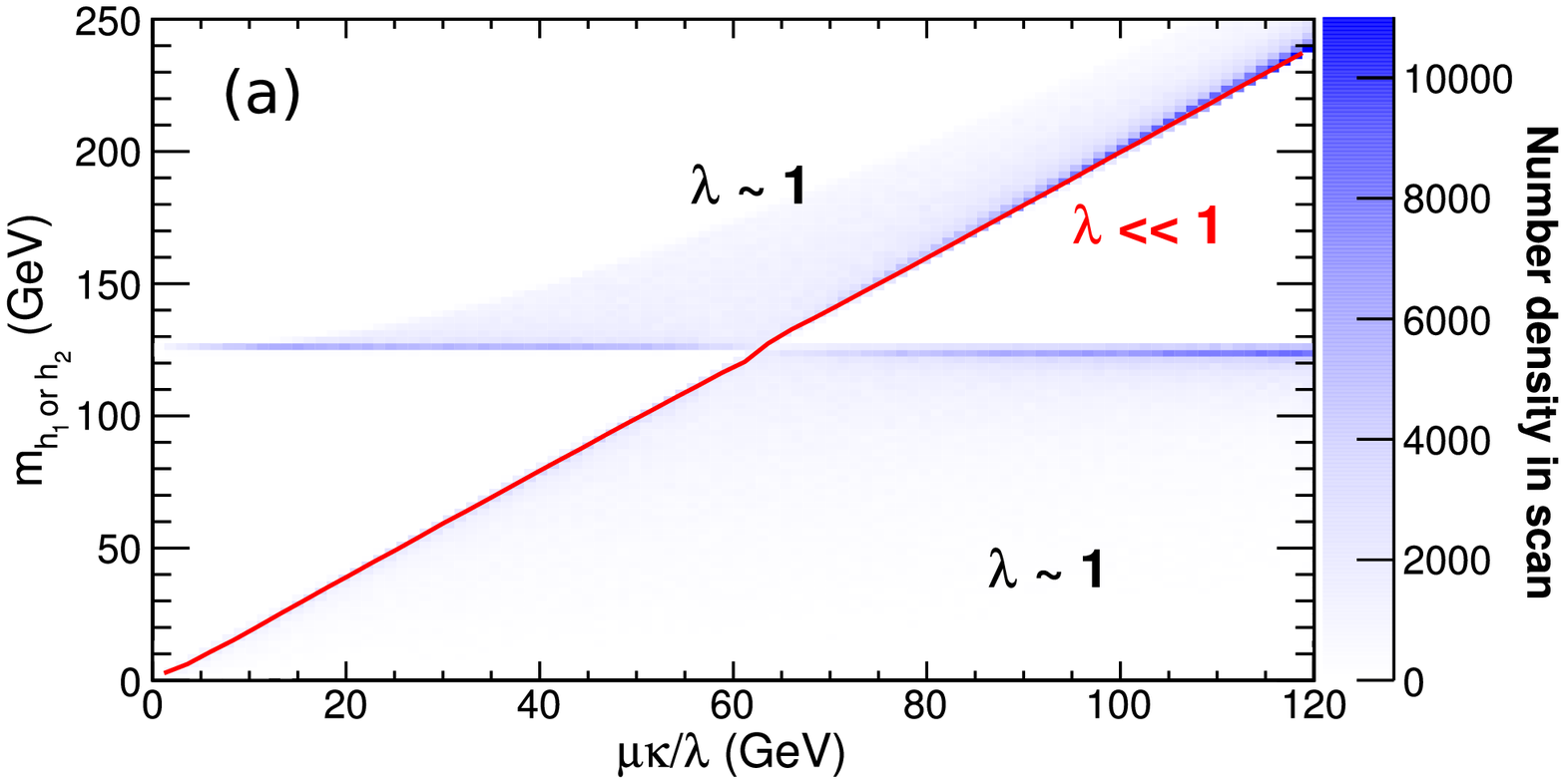}
\hfill
\includegraphics[height=0.48\linewidth,angle=90]{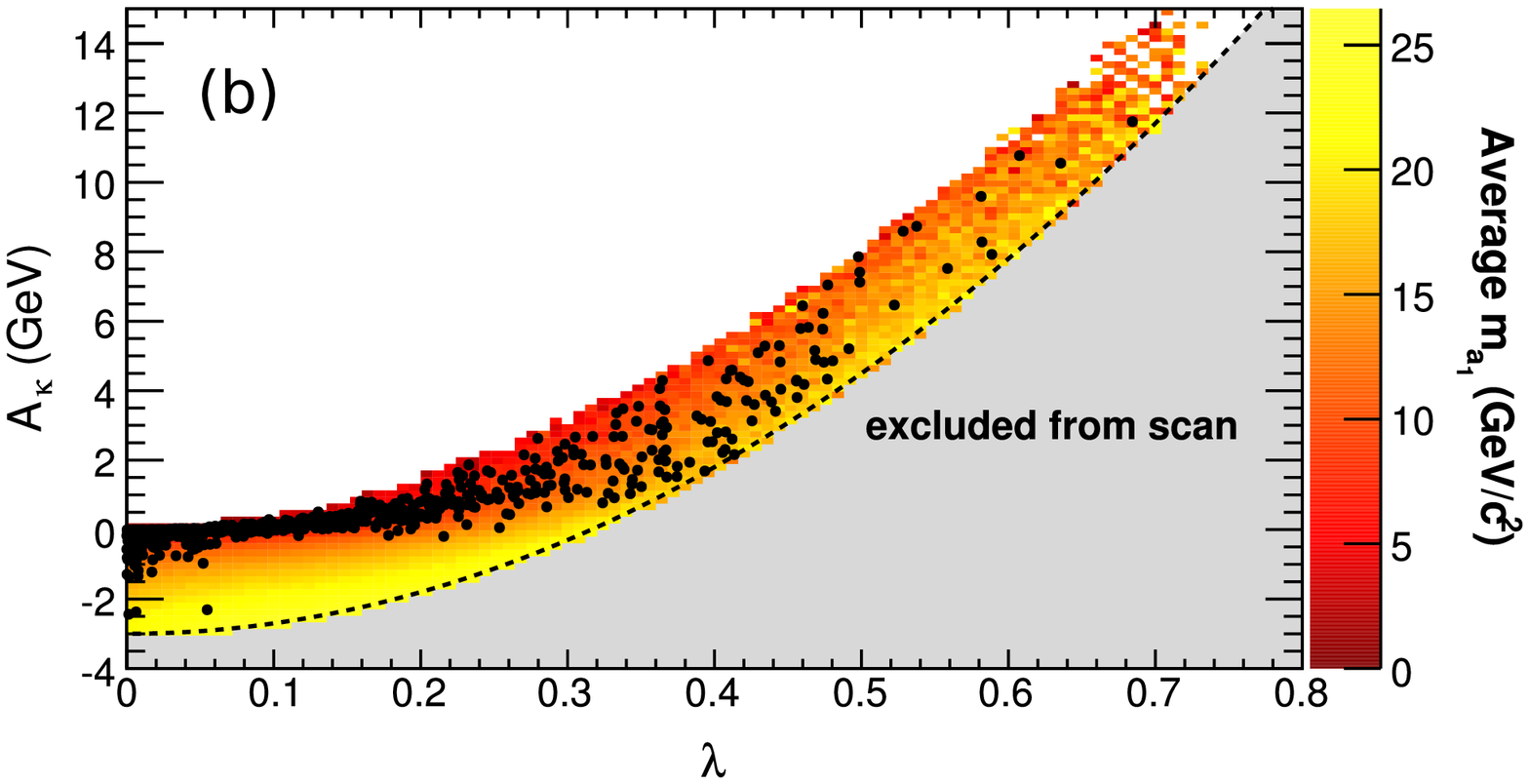}

\caption{(a):~Mass of the lightest ($h_1$) and second-lightest ($h_2$)
  CP-even Higgses as a function of $\mu\kappa/\lambda$ and $\lambda$.
  The density of generated scenarios surviving constraints is shown in
  the blue color scale, and the red line represents the single-valued
  $\lambda \ll 1$ limit (mean of $\lambda < 0.01$ scenarios).
  (b):~Mass of the lightest CP-odd Higgs ($a_1$) as a function of
  $A_\kappa$ and $\lambda$.  The color scale is the average mass in
  each bin, and filled circles are the scenarios with $m_{a_1} <
  2m_\tau$. The edge of the low $m_{a_1}$ region follows a parabolic
  curve, $(\mbox{30~GeV})\lambda^2 - A_\kappa \simeq 0$.
 \label{fig:hmass_mukoverl}}
\end{figure*}

We demonstrate that the analysis of the $4\mu$ mode has excellent
sensitivity to $h_1$ production, can be performed with early LHC data,
and requires little in terms of detector performance except reasonably
robust muon tracking and identification.  To present a realistic
analysis, we use parameters of the CMS detector at the LHC to design
event selection and to estimate background contributions.

The rest of the paper is organized as follows.  In Section~II, we
study the NMSSM parameter space in which $m_{a_1} < 2m_\tau$ and
discuss the phenomenology of the model. In Section~III, we review
constraints on the NMSSM parameter space from existing data.  In
Section~IV, we outline the proposed analysis for the 4$\mu$ mode and
evaluate its sensitivity.  Final conclusions are presented in
Section~V.

\section{NMSSM Parameter Space}

In our study we consider the simplest version of the
NMSSM~\cite{Nilles:1982dy,Frere:1983ag,Ellis:1988er,Drees:1988fc,Ellwanger:1993hn,Ellwanger:1993xa,Elliott:1993bs,Pandita:1993tg,Ellwanger:1995ru,King:1995vk,Franke:1995tc,Ellwanger:1996gw},
in which the $\mu\widehat{H_1}\widehat{H_2}$ term of the MSSM
superpotential is replaced by
\begin{equation}
\lambda \widehat{S} \widehat{H}_1 \widehat{H}_2 + \frac{\kappa}{3}  \widehat{S}^3 \mbox{,}
\label{eq:superpot} 
\end{equation}
making the superpotential scale-invariant.  In general, there are five
soft braking terms; in the ``non-universal'' case,
\begin{equation}
  m_{H_1}^2 H_1^2 + m_{H_2}^2 H_2^2 + m_{S}^2 S^2 
  + \lambda A_\lambda H_1 H_2 S +  \frac{\kappa}{3} A_\kappa S^3.
\label{eq:soft} 
\end{equation}
In the above equations, capital letters with tildes denote superfields
while symbols without tildes denote the scalar component of the
respective superfield.

Soft breaking parameters in Eq.(\ref{eq:soft}), $m_{H_1}^2$,
$m_{H_2}^2$ and $m_{S}^2$, can be expressed in terms of $M_Z$, the
ratio of the doublet Higgs VEVs, $\tan\beta$, and $\mu = \lambda s$
(where $s = \langle S \rangle$, the VEV of the singlet Higgs field)
through the three minimization equations of the Higgs potential.
Assuming that the Higgs sector is CP-conserving, the NMSSM Higgs
sector at the Electro-Weak (EW) scale is uniquely defined by
14~parameters: $\tan\beta$, the trilinear couplings in the
superpotential, $\lambda$ and $\kappa$, the corresponding soft SUSY
breaking parameters $A_\lambda$ and $A_\kappa$, the effective $\mu$
parameter $\mu = \lambda s$, the gaugino mass parameters $M_1$, $M_2$,
and $M_3$, the squark and slepton trilinear couplings $A_{t}$,
$A_{b}$, and $A_\tau$, and the squark and slepton mass parameters
$M_{f_L}$ and $M_{f_R}$.  For simplicity, we assume universality
within 3 generations for the last two parameters, leaving only 6
parameters for sfermion masses.

\subsection{Parameter Scan of the Low-$m_{a_1}$ Region of the NMSSM}

To find the parameter space for our region of interest, $m_{a_1} <
2m_\tau$, we scan the NMSSM parameters using the NMSSMTools
package~\cite{nmssmtools1,nmssmtools2,nmssmtools3}, applying all known
phenomenological and experimental constraints except the following:
the cosmological dark matter relic density measured by
WMAP~\cite{wmap}, the direct $p\bar{p} \to h_1 \to a_1 a_1 \to 4\mu$
search by the Tevatron~\cite{Abazov:2009yi}, the direct $e^+e^- \to Z
h_1$, $h_1 \to a_1 a_1$ searches by
LEP~\cite{lep1exclusion,lep2exclusion}, the direct $\Upsilon \to
\gamma a_1$ searches by CLEO~\cite{:2008hs} and
BaBar~\cite{Aubert:2009cp}, and limits set by rare $B \to K \ell^+
\ell^-$ decays~\cite{Freytsis:2009ct}.  These important constraints
are explicitly studied in our region of interest in a later section.

\begin{figure*}[htb]
\includegraphics[width=0.49\linewidth]{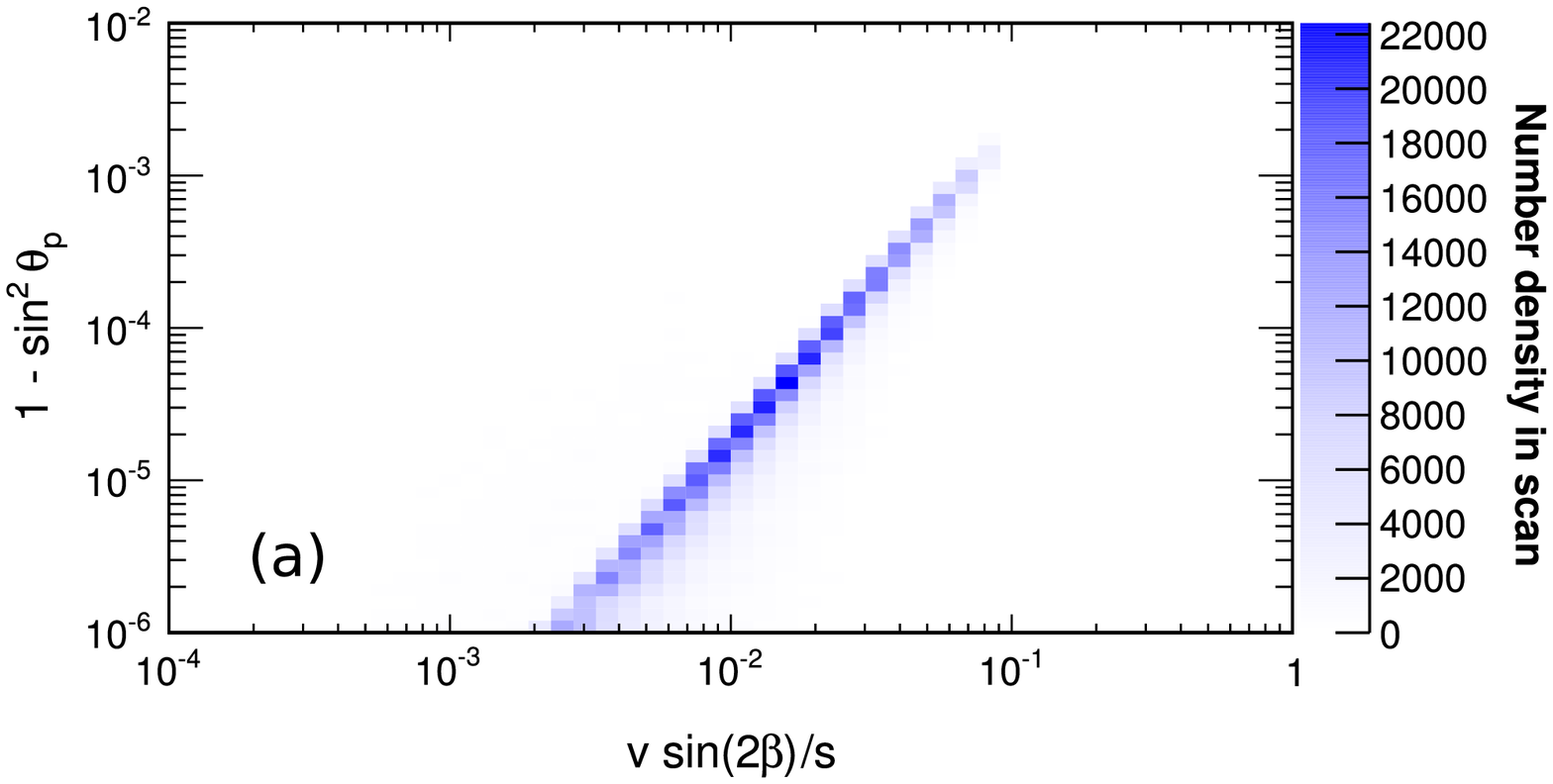}
\hfill
\includegraphics[width=0.49\linewidth]{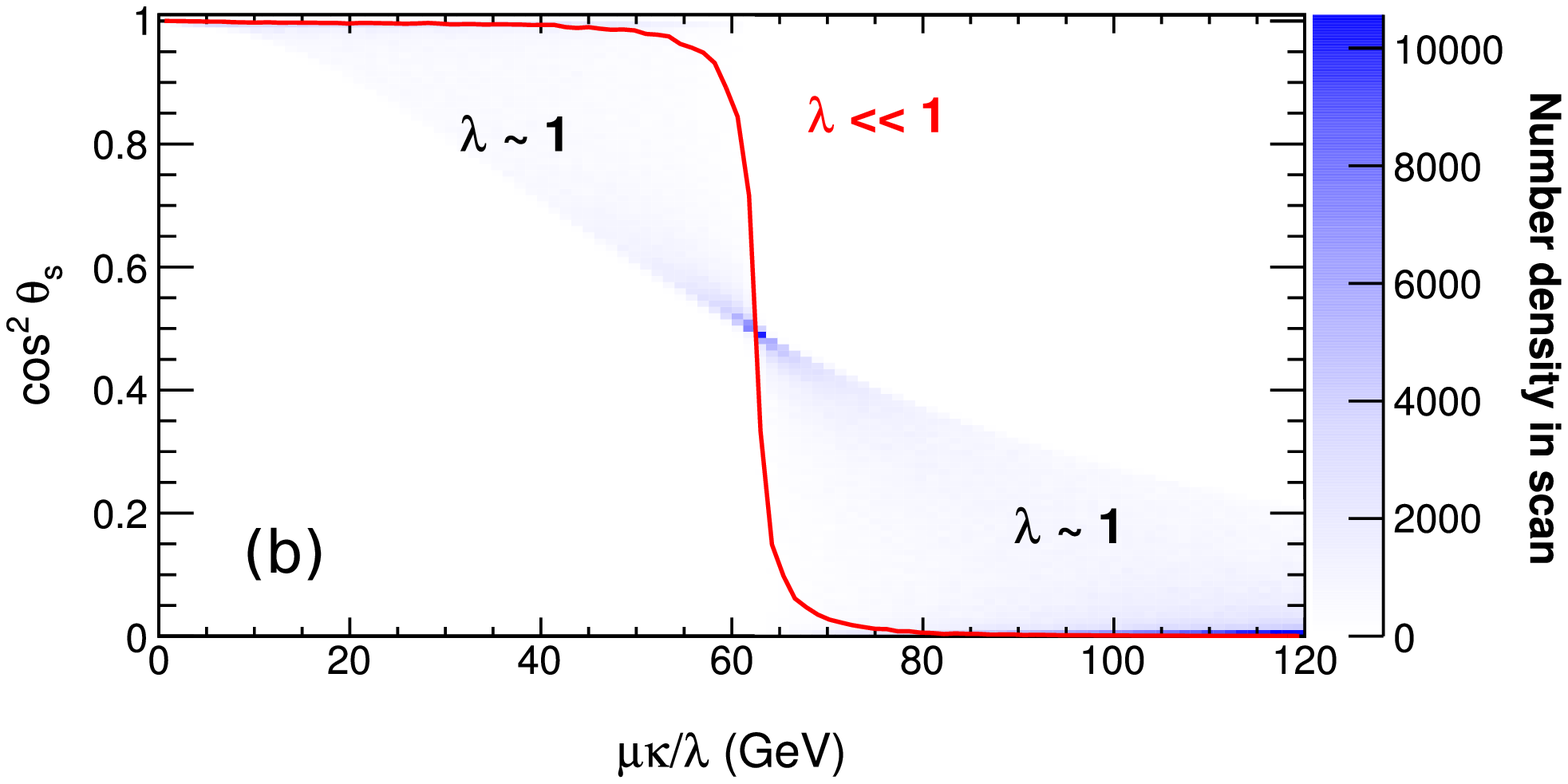}

\caption{(a):~Non-singlet fraction $1-\sin^2{\theta_P}$ of the
  lightest CP-odd Higgs boson ($a_1$) as a function of $v\sin
  2\beta/s$.  (b):~Singlet fraction $\cos^2{\theta_S}$ of the lightest
  CP-even Higgs boson ($h_1$) as a function of $\mu\kappa/\lambda$.
  (The red line is the mean of $\lambda < 0.01$ scenarios.)
\label{fig:nonsingletcp1}}
\end{figure*}

In our scan, we fix parameters entering the Higgs sector at loop-level
to $M_1/M_2/M_3 = 150/300/1000$~GeV, $A_t = A_b = A_\tau = 2.5$~TeV,
and $M_{f_L} = M_{f_R} =1$~TeV.  We then sample the NMSSM model points
uniformly in a six-dimensional space.  The first four scan parameters
are conventional, broad ranges over the probable values of $\mu$,
$\lambda$, $\tan\beta$, and $A_\lambda$:
\begin{itemize}
\item 100~GeV $<$ $\mu$ $<$ 1000~GeV
\item 0 $<$ $\lambda$ $<$ 1
\item 1.5 $<$ $\tan\beta$ $<$ 50
\item $-$1~TeV $<$ $A_\lambda$ $<$ 5~TeV
\end{itemize}

For the two remaining parameters, we identify two additional
phenomenological variables that allow a more narrow selection of the
region of interest and simplify the interpretation of our
observations.  A theoretical justification of these variables is
discussed in the next section.  The first of these two parameters,
$\mu\kappa/\lambda = \kappa s$, is selected because of its correlation
with the mass of the CP-even Higgs bosons, see
Fig.~\ref{fig:hmass_mukoverl}(a).  The corresponding range used in the
scan
\begin{itemize}
\item $0 < \mu\kappa/\lambda< 120~\mbox{GeV}$\label{eq:mh1}\\
\end{itemize}
was chosen to include two equally sized but phenomenologically
different sub-domains; in the lower one $h_1$ is light and $h_2$ is
the SM-like Higgs, and in the upper one $h_1$ is the SM-like Higgs.

The final parameter and its scan range,
\begin{itemize}
\item $0~\mbox{GeV} < (\mbox{30~GeV}) \lambda^2 - A_\kappa < 3~\mbox{GeV,}$ \label{eq:ak}
\end{itemize}
are chosen to zoom into the region of low $a_1$ masses as illustrated
in Fig.~\ref{fig:hmass_mukoverl}(b).

\subsection{Higgs Sector Spectrum and Mixings}

The CP-even and CP-odd Higgs mass matrices, $\CM_{S}$ and $\CM_{P}$,
can be written as~\cite{nmssmtools1}:
\begin{eqnarray}
\CM_{S11}^2 & = & g^2 v^2 \sin\beta^2 + \mu\tan\beta(A_\lambda + \kappa s ) 				\nonumber\\
\CM_{S22}^2 & = & g^2 v^2 \cos\beta^2 + \mu\cot\beta(A_\lambda + \kappa s )				\nonumber\\
\CM_{S33}^2 & = & \lambda A_\lambda \frac{v^2\sin 2\beta}{2s} + \kappa s(A_\kappa + 4\kappa s )   	\nonumber\\
\CM_{S12}^2 & = & (\lambda^2 - g^2/2) v^2\sin 2\beta -\lambda s (A_\lambda + \kappa s) 			\nonumber\\
\CM_{S13}^2 & = & \lambda v (2\lambda  s \cos\beta - \sin\beta (A_\lambda + 2\kappa s))			\nonumber\\ 
\CM_{S23}^2 & = & \lambda v (2\lambda  s \sin\beta - \cos\beta (A_\lambda + 2\kappa s))			
\label{eq:ms}
\end{eqnarray}
\begin{eqnarray}
\CM_{P11}^2 & = & {2\lambda s\over{\sin 2\beta}} (A_\lambda + \kappa s ) 				\nonumber\\
\CM_{P22}^2 & = & 2\lambda\kappa v^2 \sin 2\beta + \lambda A_\lambda {{v^2\sin 2\beta}\over{2s}} -3\kappa A_\kappa s \nonumber\\
\CM_{P12}^2 & = & \lambda v (A_\lambda - 2\kappa s )  
\label{eq:mp}
\end{eqnarray}

In general, $a_1$ is light in the regions of the parameter space
approaching either the Peccei-Quinn (PQ) symmetry limit ($\kappa \to
0$) or the R-symmetry (RS) limit ($A_\kappa$, $A_\lambda \to 0$).  In
both limits, $a_1$ is a massless axion, a fact which directly follows
from Eq.(\ref{eq:mp}).  It can be decomposed in terms of the weak
eigenstates $H_{uI}$, $H_{dI}$, and $S_I$ as (see
e.g.\cite{Ellwanger:2009dp}):
\begin{equation}
a_1 = \cos\theta_P A + \sin\theta_P S_I\mbox{,}
\end{equation}
where $A = \cos\beta H_{uI} + \sin\beta H_{dI}$.  In the PQ limit,
the mixing parameters $\cos\theta_P$ and $\sin\theta_P$ are
\begin{equation}
\cos \theta_P = \frac{v \sin 2\beta}{\sqrt{v^2\sin^2 2\beta + 4 s^2}}\mbox{,}
\sin \theta_P = -\frac{2s}{\sqrt{v^2\sin^2 2\beta + 4 s^2}}\mbox{.}
\label{eq:pqmix}
\end{equation}
In the RS limit, the same parameters are
\begin{equation}
\cos \theta_P = \frac{v \sin 2\beta}{\sqrt{v^2\sin^2 2\beta + s^2}}\mbox{,}
\sin \theta_P = \frac{s}{\sqrt{v^2\sin^2 2\beta + s^2}}\mbox{.}
\label{eq:rsmix}
\end{equation}

\begin{figure*}[thb]
\includegraphics[width=0.32\linewidth]{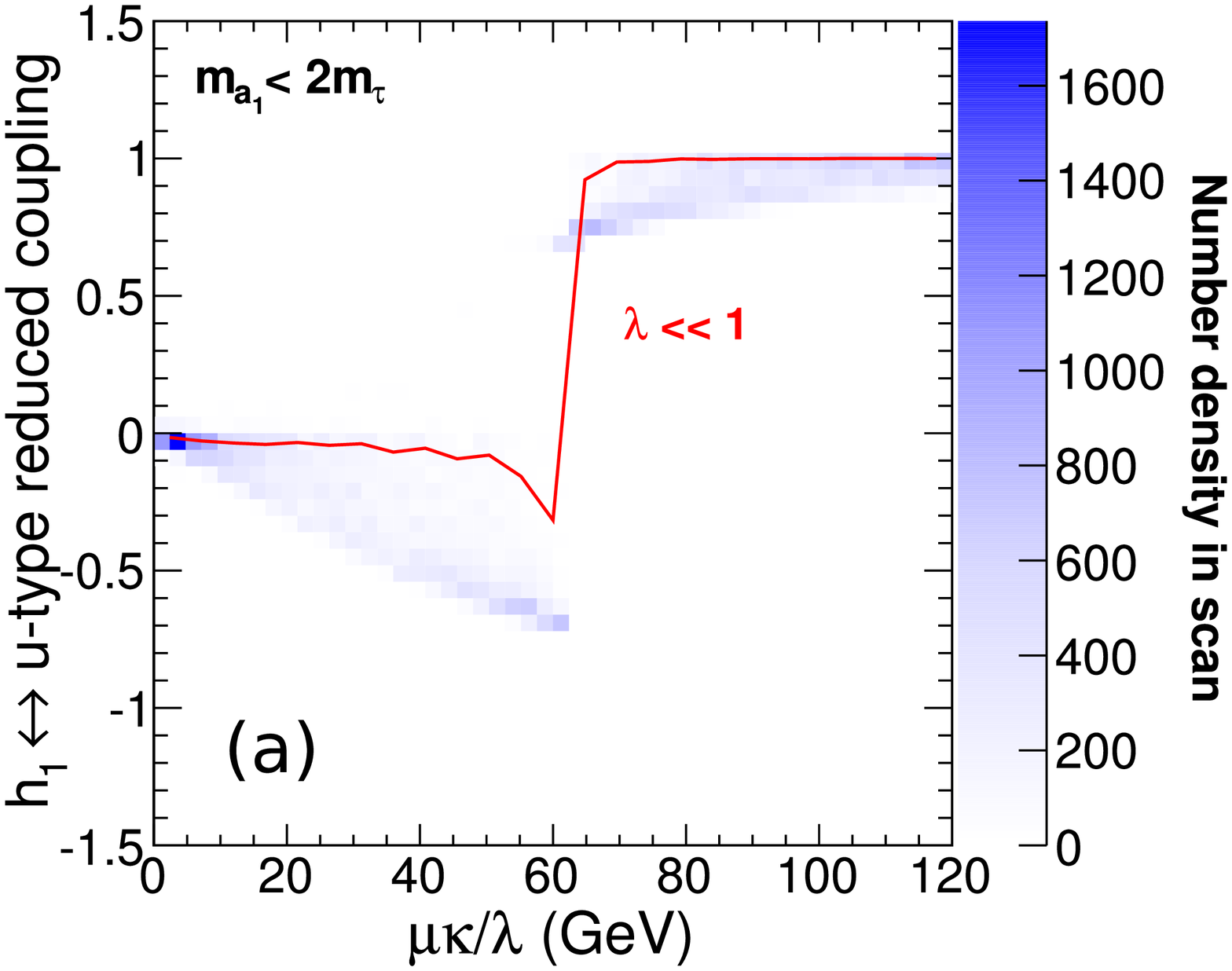}
\hfill
\includegraphics[width=0.32\linewidth]{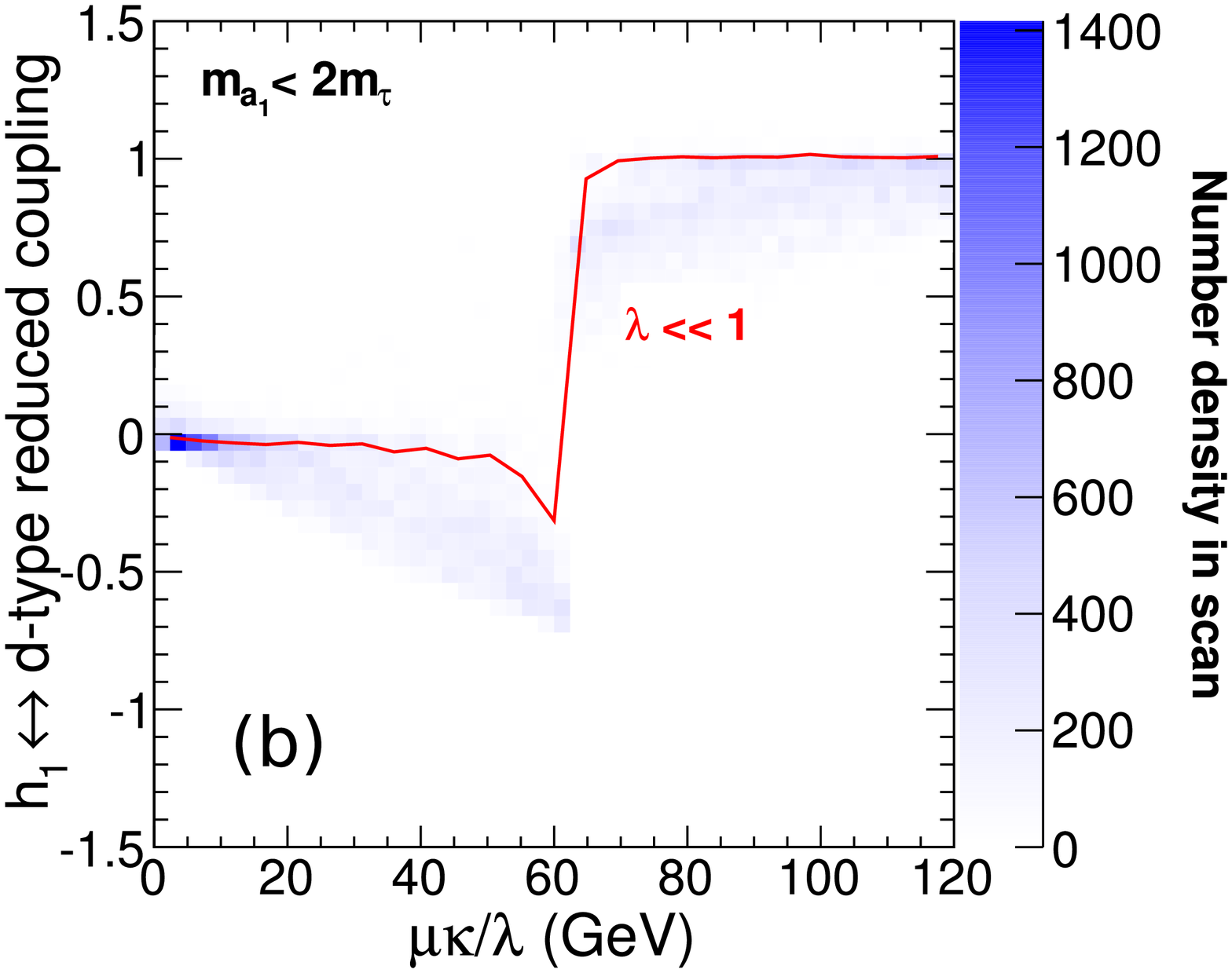}
\hfill
\includegraphics[width=0.32\linewidth]{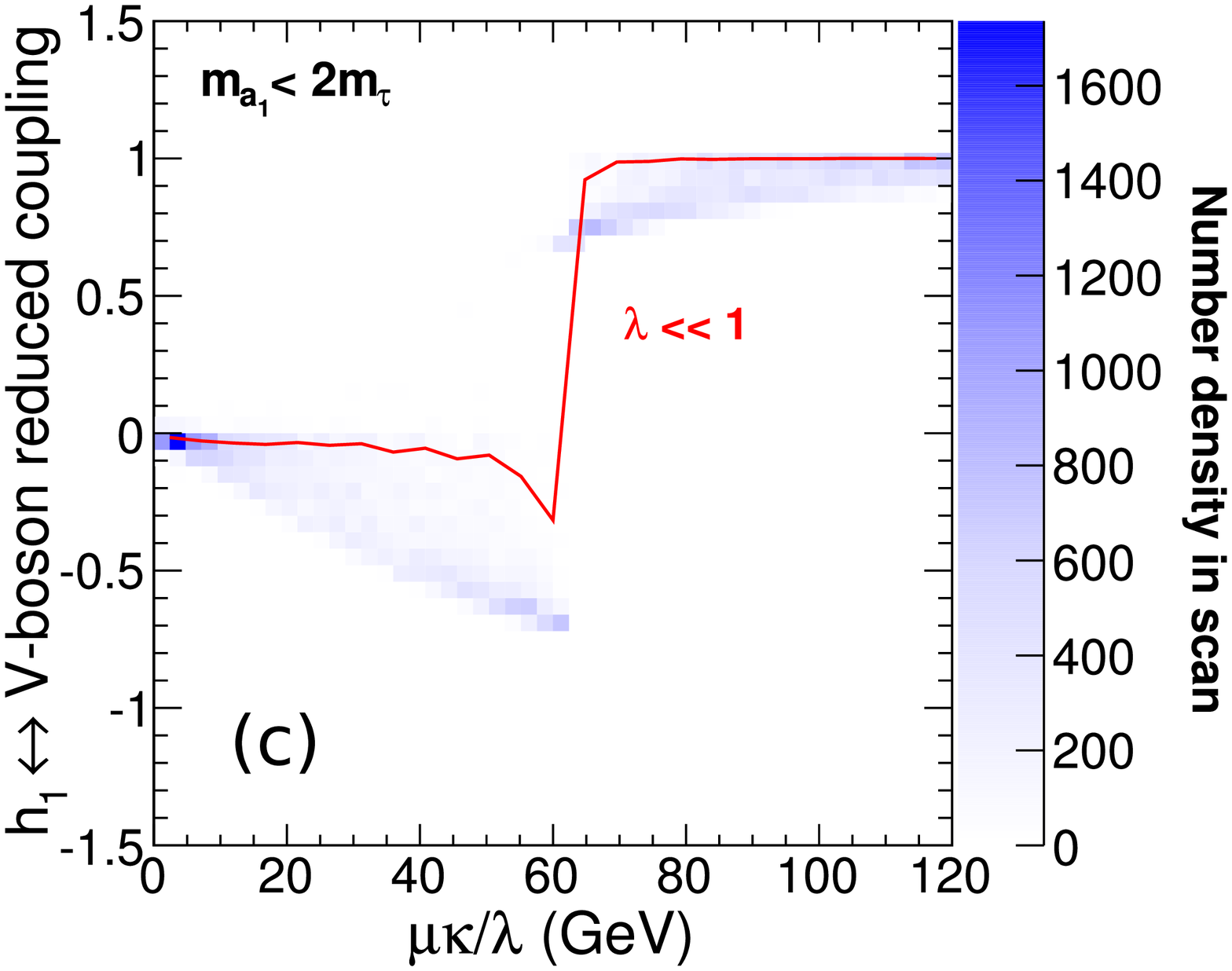}

\caption{Reduced couplings of $h_1$ to up-type quarks (a), down-type
  quarks (b), and vector bosons (c) as a function of $\mu\kappa/\lambda$, with
  the requirement that $m_{a_1} < 2m_\tau$.  (The red line is a mean
  of $\lambda < 0.01$ scenarios.) \label{fig:sm_mukoverl1}}
\end{figure*}

According to Eq.(\ref{eq:pqmix},\ref{eq:rsmix}), the non-singlet
component of $a_1$ is determined by the ratio $v \sin 2\beta / s$.
Figure~\ref{fig:nonsingletcp1}(a) shows evident correlation of the
non-singlet fraction $1 - \sin^2\theta_P$ of the lightest pseudoscalar
Higgs boson $a_1$ with $v \sin 2\beta / s$, corresponding primarily to
the PQ limit, as can be deduced from the slope of the correlation.
It also demonstrates that in the region of interest
$a_1$ is nearly a pure singlet, $1 - \sin\theta_P^2 < 1\%$, and that
the values of $v \sin 2\beta / s$ are always below $0.1$.  If we
define $\epsilon = {v \sin 2\beta / s}$, then up to
$\mathcal{O}(\epsilon)$, $\mathcal{O}(\kappa^2)$, and
$\mathcal{O}(\lambda^2)$, the CP-even Higgs mass matrix can be
re-written as
\begin{alignat}{2}
&M_S^2= \lambda s \\
\nonumber & \begin{pmatrix}
  (A_\lambda + s \kappa)\tan\beta  & - (A_\lambda + s \kappa)      & \epsilon ({\lambda s \over \sin\beta} -{A_\lambda + 2 s \kappa\over 2\cos\beta}) \\
       - (A_\lambda + s \kappa)  & (A_\lambda + s \kappa)\cot\beta  & \epsilon  ({\lambda s \over \cos\beta} -{A_\lambda + 2 s \kappa\over 2\sin\beta}) \\
\epsilon ({\lambda s \over \sin\beta} -{A_\lambda + 2 s \kappa\over 2\cos\beta}     )&   \epsilon ({\lambda s \over \cos\beta} -{A_\lambda + 2 s \kappa\over 2\sin\beta}) &
\kappa {A_\kappa+ 4 s \kappa \over \lambda}
\end{pmatrix}
\end{alignat}

One can see that $\epsilon$ also determines the mixing of singlet and
non-singlet CP-even Higgs states.  For small values of $\epsilon$ and
$A_\kappa$, characterizing the parameter space relevant to our study,
the mass of the singlet CP-even Higgs boson is determined by $2 s
\kappa = 2\mu\kappa/\lambda$.  This substantiates our earlier
observation depicted in Fig.~\ref{fig:hmass_mukoverl}(a) and the
relevance of the $\mu\kappa/\lambda$ parameter used in our scan
(Eq.\ref{eq:mh1}).  Further, in the sub-domain $\mu\kappa/\lambda <
60$~GeV, $h_1$ is light with $m_{h1} \simeq 2\mu\kappa/\lambda$ and has
a significant singlet component, particularly for smaller values of
$\lambda$ (and $A_\lambda$).  In the upper sub-domain $\mu
\kappa/\lambda > 60$~GeV, $h_1$ becomes the SM-like Higgs with
$m_{h_1} \simeq 120$~GeV while $h_2$ acquires a large singlet component
and mass $m_{h2} \simeq 2\mu\kappa/\lambda$.  This is illustrated in
Fig.~\ref{fig:nonsingletcp1}(b) showing the singlet fraction of $h_1$,
the lightest CP-even Higgs boson.

To derive $m_{a_1}$, we diagonalize the $\CM_{P}$ matrix
(Eq.~\ref{eq:mp}).  Keeping $\mathcal{O}(\kappa)$,
$\mathcal{O}(\lambda^2)$, and $\mathcal{O}(\epsilon^2)$ terms, one has
\begin{equation}
{m_{a_1}}^2 = \frac{3}{2} \, (2 \mu \kappa/\lambda)^2 \, (\zeta \lambda^2 -A_\kappa)\mbox{,}
\end{equation}
where $\zeta = \frac{3}{2} v^2\sin 2\beta/\mu$.  Because
$2\mu\kappa/\lambda$ determines the mass of the predominantly singlet
CP-even Higgs boson, $2\mu\kappa/\lambda \geq m_{h_1}$ and is always
fairly large.  Therefore, $a_1$ is light if $(\zeta \lambda^2 -
A_\kappa)$ is low, motivating the choice of the empirical parameter
$(\mbox{30~GeV}) \lambda^2 - A_\kappa$ used in the scan.  The range of
this parameter selects a region with $m_{a_1}$ between 0 and
approximately 30~GeV, avoiding most of the theoretically inaccessible
region in which $\zeta\lambda^2-A_\kappa < 0$ and therefore $m_{a_1}^2 <
0$, as shown in Fig.~\ref{fig:hmass_mukoverl}(b).

\subsection{Higgs Couplings and Decays}

The couplings of $h_1$ and $a_1$ to each other and to Standard Model
particles are determined primarily by their singlet and non-singlet
components.  While the CP-odd $a_1$ is always nearly a pure singlet
(see Fig.~\ref{fig:nonsingletcp1}(a)), the singlet fraction of $h_1$
is correlated with $\mu\kappa/\lambda$, but also depends on the
smallness of $\lambda$.  As illustrated by
Fig.~\ref{fig:nonsingletcp1}(b), for small $\lambda$, $h_1$ is nearly
a pure singlet in the $\mu\kappa/\lambda \lesssim 60$~GeV sub-region,
while in the $\mu\kappa/\lambda \gtrsim 60$~GeV domain, $h_1$ has
negligible singlet component and is essentially the SM Higgs.
Figure~\ref{fig:sm_mukoverl1} shows a strong suppression of reduced
couplings of $h_1$ to up- and down-type quarks as well as vector
bosons in the $\mu\kappa/\lambda \lesssim 60$~GeV domain.  This
suppression leads to a severe reduction in the production rates of
$h_1$ at colliders, making this scenario challenging for experimental
exploration.  Fortunately, as will be shown later, small $\lambda$
values in the low $\mu\kappa/\lambda$ region are excluded by
cosmological observations.

\begin{figure}[htb]
\includegraphics[width=0.95\linewidth]{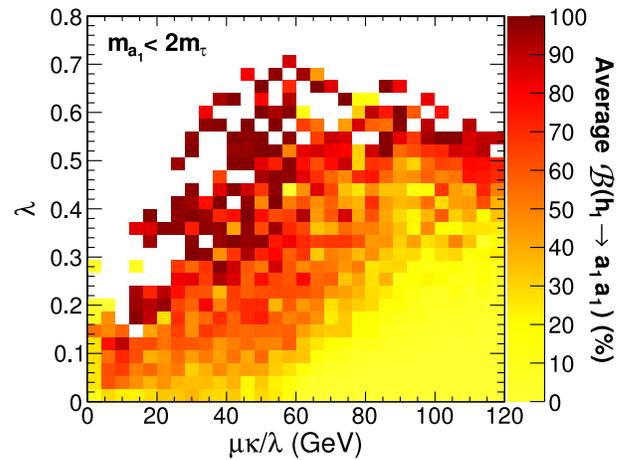}
\caption{Branching fraction of $h_1 \to a_1a_1$ in the $\lambda$
  versus $\mu\kappa/\lambda$ plane, with the requirement that $m_{a_1}
  < 2m_\tau$. \label{fig:brhaa}}
\end{figure}

Branching fractions of $h_1$ are determined by relative strength of
the $h_1$ couplings to SM particles and the $h_1 a_1 a_1$ coupling,
which is specific to the NMSSM.  Because $a_1$ has a high singlet
fraction, the singlet content of $h_1$ is directly related to the
strength of the $h_1 a_1 a_1$ coupling. If this were the only effect,
$\mathcal{B}(h_1 \to a_1 a_1)$ would have been close to 100\% in the
lower half of the $\mu\kappa/\lambda$ domain and negligible in the
upper half.  However, this coupling is also proportional to $\lambda$
(see Eq.~\ref{eq:soft}), which creates a competing effect as larger
values of $\lambda$ smear the nearly perfect separation of singlet-
and doublet-type $h_1$ in the lower and upper halves of the
$\mu\kappa/\lambda$ domain.  The overall result is illustrated in
Fig.~\ref{fig:brhaa}, showing average $\mathcal{B}(h_1 \to a_1 a_1)$
for NMSSM models with $m_{a_1} < 2m_\tau$ as a function of
$\mu\kappa/\lambda$ and $\lambda$.  It is evident that the suppression
of $h_1$ SM couplings for $\mu\kappa/\lambda < 60$~GeV makes
$\mathcal{B}(h_1 \to a_1 a_1)$ substantial as long as $\lambda$ is not
too small.  For the upper part of the $\mu\kappa/\lambda$ domain,
$\mathcal{B}(h_1 \to a_1 a_1)$ is small except for large values of
$\lambda$ where the $h_1$ singlet fraction is enhanced.

\begin{figure}[htb]
\includegraphics[height=0.95\linewidth, angle=90]{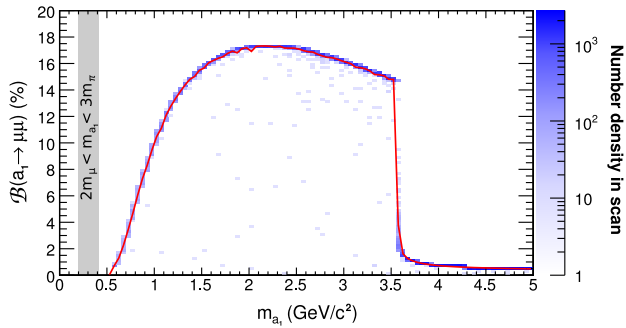}
\caption{Branching fraction of $a_1 \to \mu \mu$ for generated models
  as a function of $m_{a_1}$.  The red line is a mean of all scenarios
  as a function of $m_{a_1}$.  The threshold at 3.55~GeV/$c^2$ is $2m_\tau$.
  For $2m_\mu < m_{a_1} < 3m_\pi$ (the grey box), the branching
  fraction would be nearly 100\%. \label{fig:bramm}}
\end{figure}

As the lightest Higgs boson, $a_1$ can only decay to SM particles,
even though its coupling to SM particles is strongly suppressed due to
its nearly singlet nature.  One should also notice that $a_1$
couplings to down-type fermions are proportional to $\tan\beta$ while
its couplings to up-type fermions are suppressed as $1/\tan\beta$.
Therefore, $a_1$ branching fractions follow the standard mass
hierarchy of open decay channels to down-type fermions.
Figure~\ref{fig:bramm} shows the the branching fraction for $a_1 \to
\mu\mu$ as obtained using NMSSMTools package.  For $m_{a_1} <
2m_\tau$, the $a_1 \to \mu\mu$ channel becomes significant, making an
analysis in the four-muon mode viable for experimental searches.

\begin{figure*}[htb]
\includegraphics[width=0.87\linewidth]{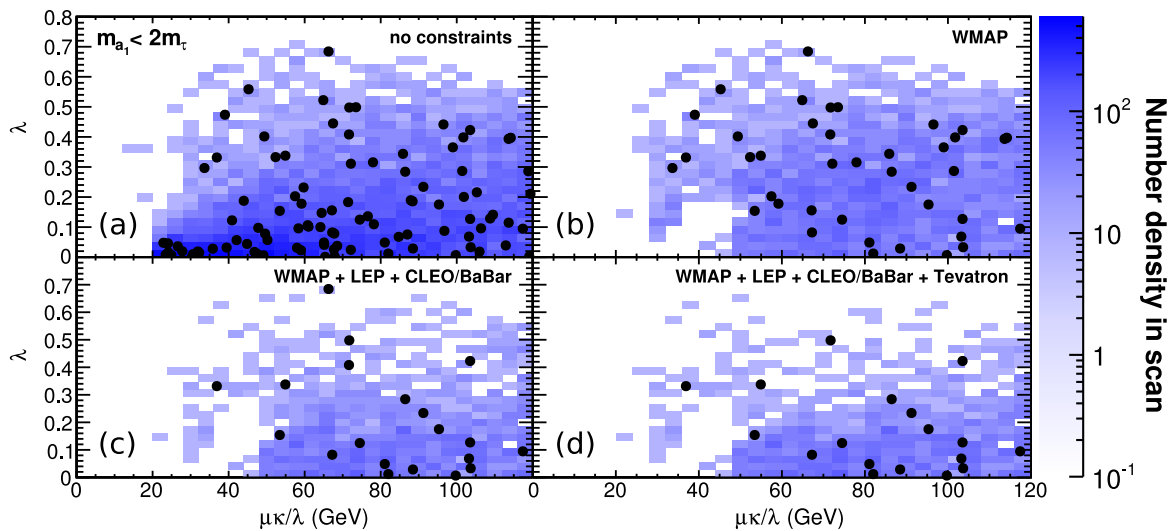}
\caption{Sampled points with $m_a < 2m_\tau$ and experimental constraints successively applied 
in the $\lambda$ vs.\ $\mu\kappa/\lambda$ plane. The low energy $e^+e^-$ data (CLEO and
BaBar) have essentially no impact on the allowed parameter space. Color scale is number density 
and filled points are 100 models (before application of experimental 
constraints). \label{fig:exclusion_params}}
\end{figure*}
\begin{figure*}[hbt]
\includegraphics[width=0.87\linewidth]{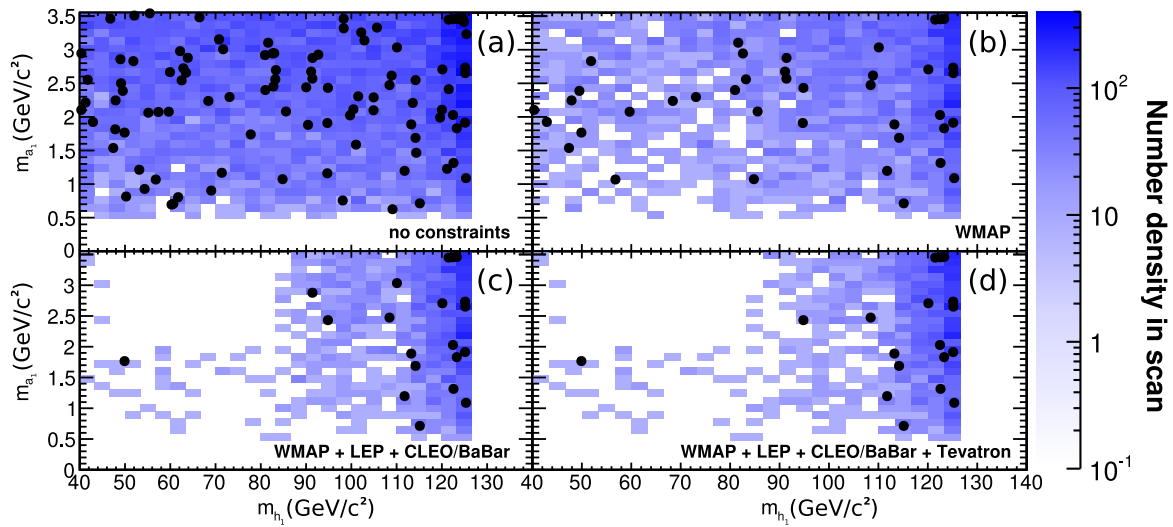}
\caption{Sampled points with $m_a < 2m_\tau$ and experimental constraints successively applied 
similar to Fig.~\ref{fig:exclusion_params} but in the $m_a$ vs. $m_h$ plane.  The low 
energy $e^+e^-$ data (CLEO and BaBar) have essentially no impact on the allowed parameter space. 
Color scale is number density and filled points are 100 models (before application of experimental 
constraints). \label{fig:exclusion_mass}}
\end{figure*}

It is important to note that the NMSSMTools calculation of
$\mathcal{B}(a_1 \to \mu \mu)$ shown in Fig.~\ref{fig:bramm} does not
include hadronization effects important in the region $m_{a_1} <
1$~GeV/$c^2$, and therefore is not reliable.  One should also notice
that for $2m_\mu < m_{a_1} < 3m_\pi$, $\mathcal{B}(a_1 \to \mu\mu)$ is expected
to be about 100\% because $q\bar{q}$ and $gg$ decays are prohibited by
hadronization and spin effects, and because $\gamma\gamma$ is small.
Since the status of $\mathcal{B}(a_1 \to \mu \mu)$ in NMSSMTools for
$m_{a_1} < 0.5$~GeV/$c^2$ is not well-established and requires further
development, we present our results only for the $m_{a_1} > 0.5$~GeV
region.  We would like to notice that $\mathcal{B}(a_1 \to \mu \mu)$
is model-dependent in general and can be somewhat different in, for
example, the Little Higgs model~\cite{LHiggsADecays} used by the $D\O$
collaboration in Ref.~\cite{Abazov:2009yi}.  Therefore, we present our
results as limits on production cross-section times branching ratios
for the $4\mu$ signature as a model-independent limit for a given
topology and signature under study.

\section{Current Experimental Constraints}

Existing experimental data restrict the NMSSM parameter space for the
scenario studied here.  In the following, we discuss experimental
measurements relevant to this scenario and evaluate their impact in
restricting the allowed parameter space for models with low $m_{a_1}$.

\subsection{Cosmological Constraints}

The lightest NMSSM neutralino is a candidate for Cold Dark Matter
(CDM).  The WMAP measurement of the CDM relic density therefore serves
as an important constraint on the allowed NMSSM parameter space.  In
our scan, we used the MicrOmegas package~\cite{micrOmegas} linked to
NMSSMTools to calculate $\Omega_{\mbox{\scriptsize NMSSM}}h^2$ and
determine if a particular model is consistent with the experimental
data.  We considered a model to be consistent with the CDM measurement
if $\Omega_{\mbox{\scriptsize NMSSM}}h^2 \le 0.1131 + 2\times0.0034$,
corresponding to the 95\% upper limit obtained using the latest WMAP
5-year dataset~\cite{wmap}.  The NMSSM neutralino relic density need
not account for all CDM observed by WMAP, but it cannot exceed it.

To illustrate the effect of the WMAP constraints,
Fig~\ref{fig:exclusion_params}(a) shows the density of generated NMSSM
models in the $\lambda$ versus $\mu\kappa/\lambda$ plane under the
constraint that $m_{a_1} < 2m_\tau$.  Models that were determined to
be consistent with the WMAP data are shown in
Fig.~\ref{fig:exclusion_params}(b).  The comparison shows that the
WMAP bound excludes the region of small $\mu\kappa/\lambda$ and
$\lambda$.  In that region, the lightest neutralino is light and
weakly interacts with SM particles, suppressing the neutralino
annihilation rate and enhancing the neutralino relic density to
unacceptably large values.  Figures~\ref{fig:exclusion_mass}(a) and
(b) make the same comparison but in the $m_{a_1}$ versus $m_{h_1}$
plane.

\subsection{Constraints from Direct Searches at Colliders}

Several searches for $h_1 \to a_1 a_1$ have been performed at collider
experiments, with the strongest impact on the allowed NMSSM models
coming from LEP-II data~\cite{lep2exclusion}.  Although the singlet
component of $h_1$ at low $\mu\kappa/\lambda$ and $\lambda$ (and
correspondingly low $m_{h_1}$) would severely suppress $h_1$
production in $e^+e^- \to h_1 Z$, these extreme scenarios are excluded
by the WMAP data.  LEP limits~\cite{lep2exclusion} on NMSSM models are
inferred from $h_1 \to a_1 a_1$, $a_1 \to$ pairs of charm, gluon, and
$\tau$ jets; $a_1 \to \mu\mu$ limits were not quoted.  The LEP-II
upper limit on $e^+ e^- \to h_1 Z$ with $h_1 \to a_1 a_1$ excludes
models that predict $m_{h_1}$ within the kinematic limits, $45 <
m_{h_1} < 86$~GeV/$c^2$, and $m_{a_1}$ in the region of significant
detector efficiency, $m_{a_1} > 2$~GeV/$c^2$.

In addition to LEP data, there were direct searches for $\Upsilon \to
\gamma a_1$ by CLEO and BaBar at low energy $e^+e^-$
colliders~\cite{cleo-low-ma,babar-low-ma}.  Neither of these searches
significantly constrain the NMSSM models with low $m_{a_1}$ because
$a_1$ has a high singlet component, and thus negligible $bba_1$
coupling (see Fig.~\ref{fig:nonsingletcp1}(a)), for all sampled
parameter values.  Because CLEO and BaBar results have negligible
effect on the allowed parameter space,
Figs.~\ref{fig:exclusion_params}(c) and~\ref{fig:exclusion_mass}(c)
show combined LEP+CLEO+BaBar constraints, but the reader is reminded
that only LEP constraints are relevant.

\begin{figure*}[htb]
\includegraphics[width=0.48\linewidth]{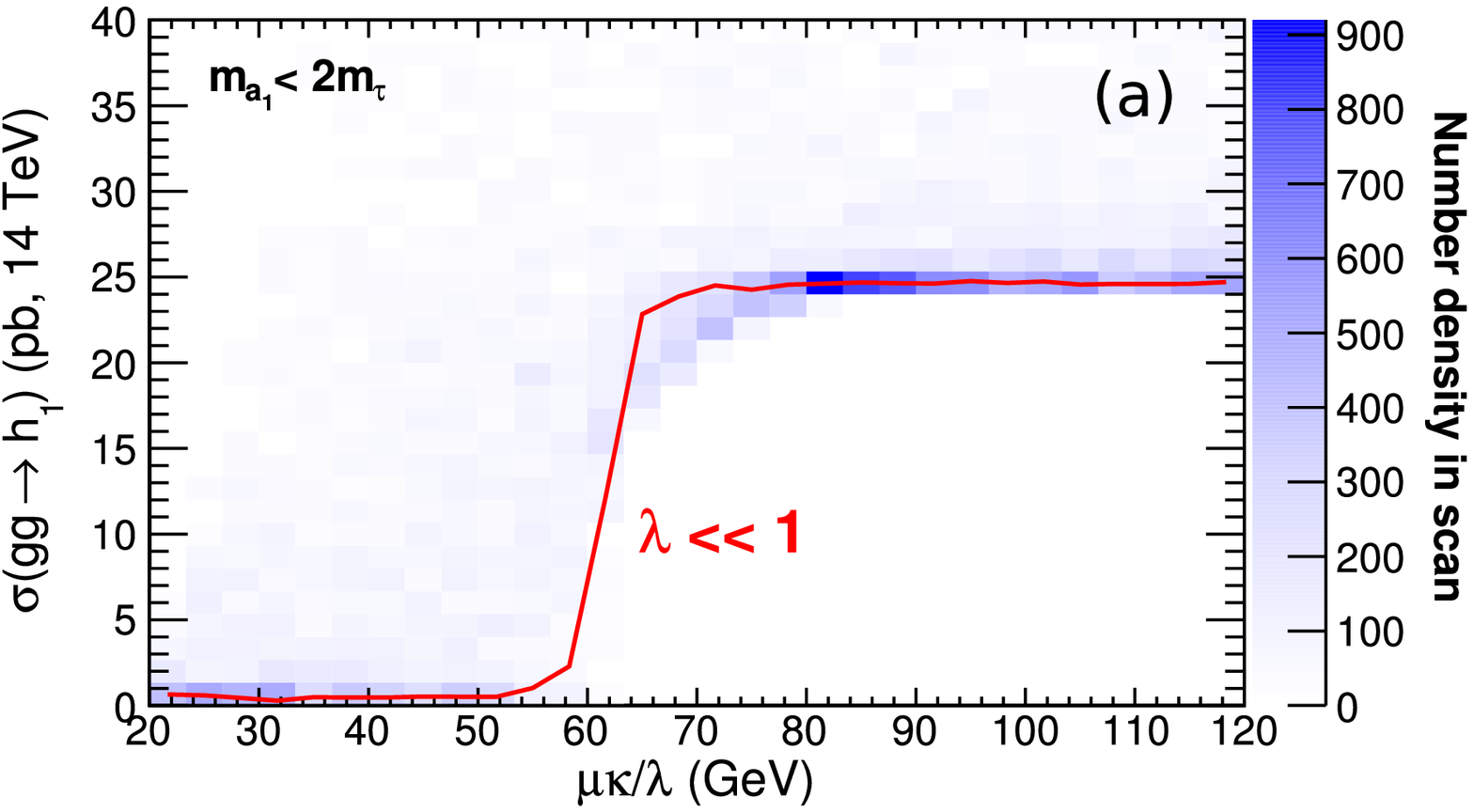}
\hfill
\includegraphics[width=0.48\linewidth]{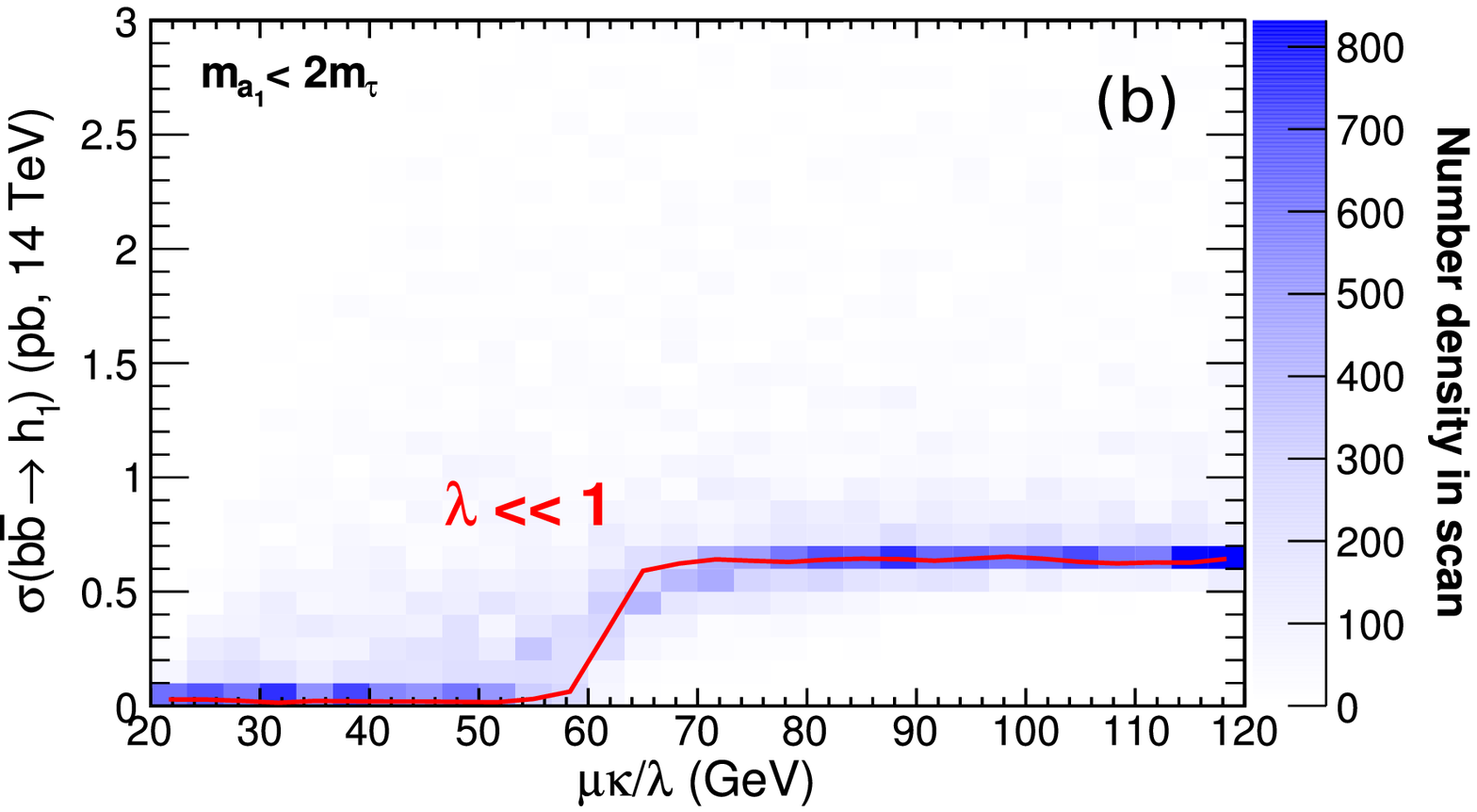}

\caption{Contributions to the $pp \to h_1$ production cross-section at
  $\sqrt{s}=14$~TeV as a function of $\mu\kappa/\lambda$ and
  $\lambda$, from $gg$ (a) and $b\bar{b}$ (b), with the requirement
  that $m_{a_1} < 2m_\tau$.  (The red line is a mean of $\lambda <
  0.01$ scenarios.)
  \label{fig:sm_mukoverl2}}
\end{figure*}

The flat $\ell^+\ell^-$ distribution of rare $B \to K \ell^+\ell^-$
decays~\cite{:2008sk} could potentially set a limit on the parameter
space under study through the bound on $\mathcal{B}(B \to K a_1)
\times \mathcal{B}(a_1 \to \mu\mu)$. However, this limit does not
actually bound the region of our interest because the coupling of
$a_1$ to up-type and down-type quarks in $b \to s a_1$ penguin
diagrams is suppressed due the highly singlet nature of $a_1$.  For
$f_a = \tan\theta_P \, v \sin(2\beta)/2$, the quantity $f_a \tan^2\beta$
is above 100~TeV in our scan while the charged Higgs mass is typically
above 200~GeV/$c^2$.  Thus, the bound set in Fig.~2 of
Ref.~\cite{Freytsis:2009ct} is not relevant to the parameter space of our
study.

The results of a search~\cite{d0-low-ma} for the NMSSM with a low mass
$a_1$ at the Tevatron was recently published by the $D\O$
collaboration in the channel $h_1 \to a_1 a_1 \to 4\mu$.  With no
excess of data over the SM expectations, the paper quotes 95\%
C.L.\ upper limits for the cross-section of this process.  To
interpret the $D\O$ result in terms of constraints on allowed NMSSM
models in our scan, we calculate the NLO production cross-section for
$p\bar{p} \to h_1$ in the NMSSM using the SM NLO calculations for $gg
\to H_{SM}$~\cite{Spira:1995rr} and $b\bar{b} \to H_{SM}$ with
QCD-improved (running) Yukawa couplings~\cite{Balazs:1998sb},
corrected for differences in coupling between the SM and the NMSSM
using NMSSMTools:
\begin{align}
\sigma(gg\to h_1)=\sigma(gg\to H_{SM})\frac{\Gamma(h_1\to gg)}{\Gamma(H_{SM}\to gg)} \label{eq:ggcross_section} \\
\nonumber =\sigma(gg\to H_{SM})\frac{Br(h_1\to gg)\Gamma^{\mbox{\scriptsize tot}}(h_1)}{\Gamma(H_{SM}\to gg)} \\
\sigma(b\bar{b}\to h_1)=\sigma(b\bar{b}\to H_{SM})
\left(\frac{Y_{bbh_1}}{Y_{bbH_{SM}}}\right)^2 \label{eq:bbcross_section} 
\end{align}
where $\sigma(gg \to H_{SM})$ and $\Gamma(H_{SM} \to gg)$ are
calculated using HIGLU~\cite{higlu}, while $\mathcal{B}(h_1 \to gg)$,
$\Gamma^{\mbox{\scriptsize tot}}(h_1)$, and the ratio of Yukawa
couplings $Y_{bbh_1}/Y_{bbH_{SM}}$ are obtained using NMSSMTools.  For
$\mu\kappa/\lambda < 60$~GeV (non-SM $h_1$ lighter than
120~GeV/$c^2$), the cross-section is strongly suppressed even compared
to the SM for low $m_{a_1}$ because $h_1$ has a large singlet fraction
and weakly couples to SM particles (see Fig.~\ref{fig:sm_mukoverl1}).
For larger $\mu\kappa/\lambda$, the lightest CP-even Higgs $h_1$
becomes the SM-like Higgs and has a small $h_1 \to a_1 a_1$ branching
fraction.

The $D\O$ paper~\cite{d0-low-ma} quotes 95\% C.L.\ limits on
$\sigma(p\bar{p} \to h_1) \times B(h_1 \to a_1 a_1 \to 4\mu)$ for
several choices of $m_{a_1}$ with $m_{h_1} = 100$~GeV/c$^2$.  To
determine if a particular model in our scan is excluded by these data,
we linearly interpolate the published cross-section limits for values
of $m_{a_1}$ between the points in~\cite{d0-low-ma}.  To obtain the
experimental cross-section limits as a function of $m_{h_1}$, we need
to correct for variations in the experimental acceptance.  We obtain
those limits by taking the analysis acceptance to be linear as a
function of $m_{h_1}$ ``increasing by $\sim$10\% when $m_{h_1}$
increases from 80 to 150~GeV/$c^2$''~\cite{d0-low-ma} and matching it
to the full analysis acceptance given at $m_{h_1} = 100$~GeV/$c^2$.
We then calculate the production cross-section and branching fractions
for the model points and compare them to the values we derive
from~\cite{d0-low-ma}.  Figures~\ref{fig:exclusion_params}(d)
and~\ref{fig:exclusion_mass}(d) show the density of NMSSM models
surviving WMAP, LEP and Tevatron constraints.  Because of the
suppression in production rate at low $\mu\kappa/\lambda$ and small
$\mathcal{B}(h_1 \to a_1 a_1)$ at high $\mu\kappa/\lambda$, the
Tevatron search has only a limited impact on the allowed NMSSM
parameter space, mainly excluding models with high $\lambda$.  A
significant improvement in Tevatron reach for the NMSSM would require
a large increase in integrated luminosity, thus requiring the LHC to
make a definitive discovery or exclusion of NMSSM models with low
$m_{a_1}$.

\section{A Dedicated Search for the Low-$m_{a_1}$ NMSSM at the LHC}

\begin{figure*}[htbp]
\begin{center}
\includegraphics[width=0.48\linewidth]{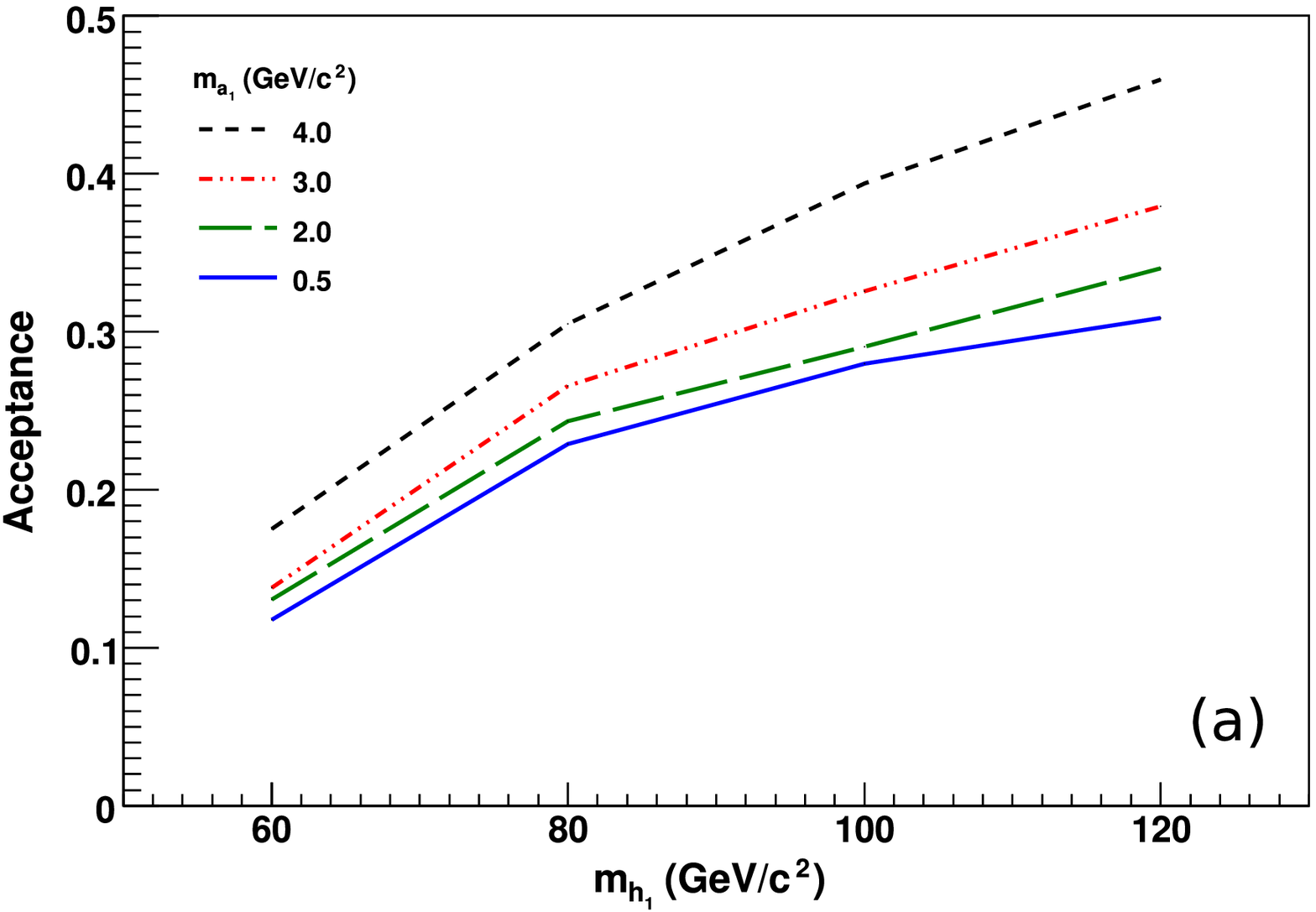}
\hfill
\includegraphics[width=0.48\linewidth]{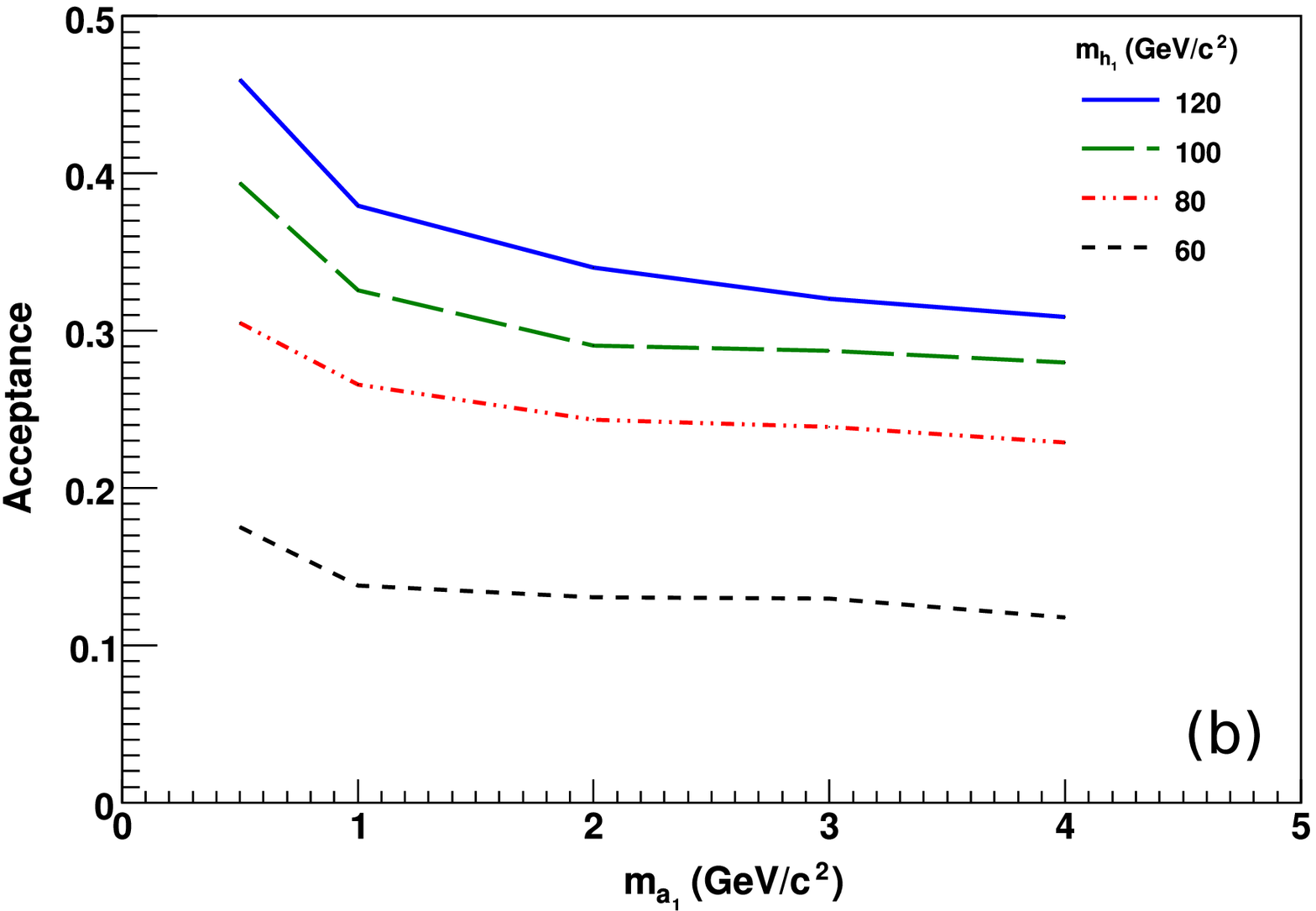}

\caption{(a):~Acceptance as a function of $m_{a_1}$ for fixed
  $m_{h_1}$.  (b):~Acceptance as a function of $m_{h_1}$ for fixed
  $m_{a_1}$.}
\label{signal_acceptance}
\end{center}
\end{figure*}

Since $a_1$ is only weakly coupled to SM particles, it can only be
produced at the LHC via decays of the lightest scalar Higgs $h_1 \to
a_1 a_1$.  The main characteristic of such signal at the LHC is two
back-to-back (in $\phi$) di-muon pairs of spatially nearby muons.  The
reconstructed di-muon pairs should have invariant masses consistent
with one another, and their invariant masses also serve as a direct
measurement of $m_{a_1}$.  Additionally, the $4\mu$ invariant mass
distribution should have a narrow peak corresponding to the $m_{h_1}$
mass.  We use these striking features to design an analysis suitable
for early LHC running.

The four-muon final state considered in this analysis has relatively
low experimental backgrounds.  Therefore, instead of using the Vector
Boson Fusion (VBF) production process chosen in the proposed NMSSM
searches targeting the $m_{a_1} > 2m_\tau$ region~\cite{nmssm-ph7}, we
focus on the largest Higgs production modes at the LHC, $gg \to h_1$
and $b\bar{b} \to h_1$.  We calculate the NLO cross-section for $pp
\to h_1$ for the NMSSM by rescaling the LHC SM NLO
calculations~\cite{Spira:1995rr,Balazs:1998sb} to correct for
differences in couplings between the SM and NMSSM
(Eqs.~\ref{eq:ggcross_section} and~\ref{eq:bbcross_section}).  Like
the Tevatron case, the cross-section is strongly suppressed compared
to the SM if $h_1$ has a large singlet fraction.
Figure~\ref{fig:sm_mukoverl2} shows the production cross-section for
14~TeV $pp \to h_1+X$ as a function of $\mu\kappa/\lambda$.  While
this suppression makes the analysis challenging even at the LHC, the
constraints arising from the WMAP relic density measurement exclude
models with very low values of $\lambda$, so the allowed models have
small but non-negligible production cross-section.

\subsection{Analysis Selections}

We use Pythia to generate signal event templates with $m_{h_1}$ in the
range from 70 to 140~GeV/$c^2$ and $m_{a_1}$ in the range from 0.5 to
4~GeV/$c^2$.  We chose the CMS detector as a benchmark for modeling a
realistic experimental environment, with parameters described in
Ref.~\cite{cms-tdr}.  The important parameters for this analysis are
muon momentum resolution, the minimum muon momentum needed to reach
the muon system, geometric acceptance, and the average muon
reconstruction efficiencies.  Because of the large number of
reconstructed muons in the event, we take the trigger efficiency to be
100\%.

The analysis starts by requiring at least four muon candidates with
transverse momentum $p_T > 5$~GeV/$c$ and pseudorapidity $|\eta| <
2.4$ to ensure high and reliable reconstruction efficiency.  Of the
four muon candidates, at least one must have $p_T > 20$~GeV/$c$ to
suppress major backgrounds and to satisfy trigger requirements.  Each
event is required to have at least two positively charged and two
negatively charged muon candidates.  For the surviving events, we
define quadruplets of candidates, pairing the candidates by charge and
sorting them into two di-muon pairs by minimizing the quantity
$(\Delta R(\mu_i,\mu_j)^2 + \Delta R (\mu_k,\mu_l)^2)$, where $\Delta
R^2 = \Delta \eta^2 + \Delta \phi^2$.  Muon quadruplets for which
$\Delta R > 0.5$ in either of the pairs are discarded as inconsistent
with the signal topology.  Acceptance for the selections listed above
is shown in Fig.~\ref{signal_acceptance} for several representative
values of $m_{h_1}$ and $m_{a_1}$.

The requirement of four sufficiently energetic muons in the event
dramatically reduces contributions of potential backgrounds for this
analysis.  After acceptance selections, the dominant background is due
to the QCD multijet production where muons originate from heavy-flavor
resonances, heavy-flavor quark decays, or from $\pi/K$
decays-in-flight.  We use Pythia to estimate the QCD multijet
background, and obtain approximately 2.6~events/pb$^{-1}$
(approximately half containing at least one decay-in-flight).  Using
CalcHEP~\cite{calchep} to estimate $pp \to 4\ell + X$ electroweak
backgrounds, we obtain 0.04~events/pb$^{-1}$.  Direct $J/\psi$
production is found by Pythia to be completely negligible.  Other SM
backgrounds (top, W+jets) are negligible in the region of interest of
this analysis.

\begin{figure*}[htb]
\begin{center}
\includegraphics[width=0.48\linewidth]{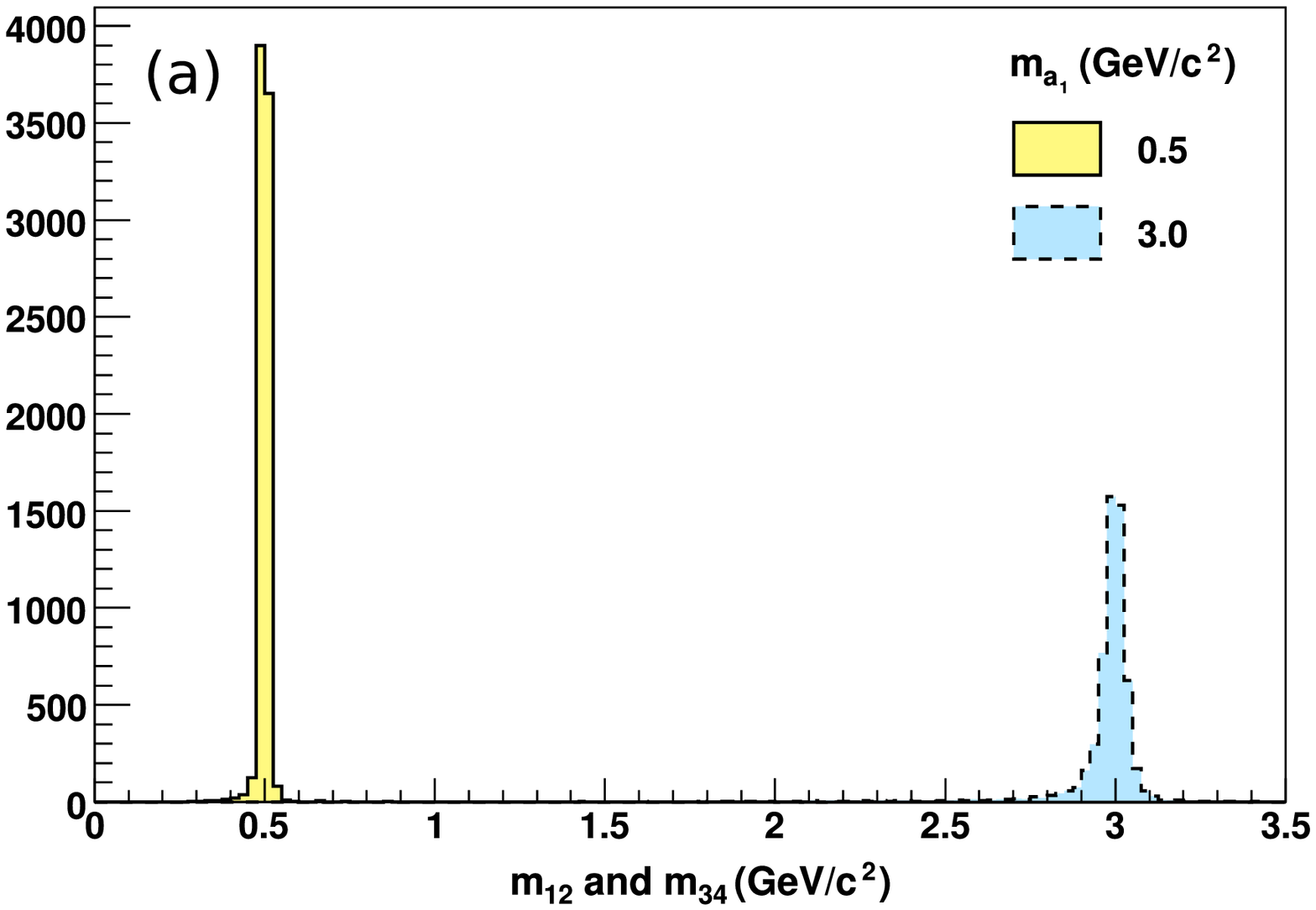}
\hfill
\includegraphics[width=0.48\linewidth]{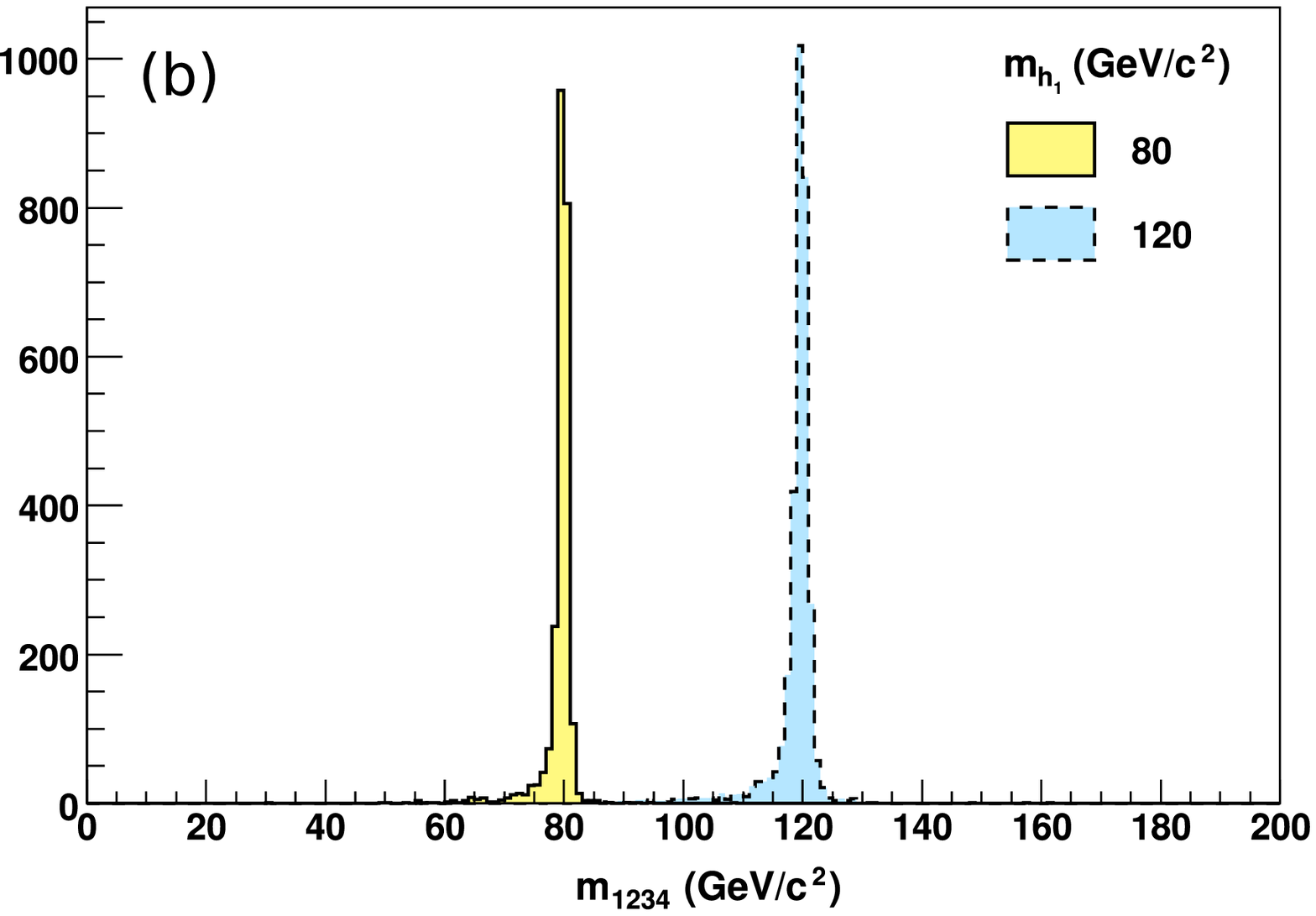}

\caption{(a):~Invariant mass of reconstructed muon pairs for $m_{a_1}
  = 0.5$ and 3~GeV/c$^2$ (in both cases $m_{h_1}$ = 100~GeV/c$^2$).
  (b):~Invariant mass of four reconstructed muons for $m_{h_1} = 80$ and
  120~GeV/c$^2$ (in both cases $m_{a_1}$ = 3.0~GeV/$c^2$).}
\label{muon_pairs_masses_invariant_mass}
\end{center}
\end{figure*}

The backgrounds are further reduced by requiring the kinematics to be
consistent with the expected signal siganture.  We calculate the
invariant mass of each of di-muon pair, $m_{12}$ and $m_{34}$, as well
as the invariant mass of all four muons, denoted as
$m_{1234}$. Figure~\ref{muon_pairs_masses_invariant_mass}(a) shows the
invariant mass of the muon pairs passing all selections in signal
events for two choices of $m_{h_1}$ and
$m_{a_1}$. Figure~\ref{muon_pairs_masses_invariant_mass}(b) shows the
distribution of $m_{1234}$ for two benchmark points.  To focus on the
region of interest, we require $m_{1234} > 60$~GeV/$c^2$, $m_{12}$,
$m_{34} < 4$~GeV/$c^2$, which reduces the QCD background to
0.4~events/pb$^{-1}$.

To ensure the compatibility of the measured invariant masses of the
two di-muon pairs, one could require $|m_{12} - m_{34}| < 0.08 +
0.005\times(m_{12} + m_{34})$. Such cut would require the two pair
masses to be consistent with each other and would take into account
the widening of absolute resolution in the reconstructed di-muon mass
as a function of mass.  If applied, the only background that still may
be not completely negligible is the QCD multi-jet production, for
which we conservatively estimate the upper bound to be
0.02~events/pb$^{-1}$.  However, instead of applying this selection
explicitly, a better approach would be to fit the data in the 3D space
$(m_{12}, m_{34}, m_{1234})$, taking into account kinematic properties
of the signal events.  This approach maximizes the signal acceptance
and therefore the statistical power of the analysis.  It is also
convenient from an experimental standpoint as the background events
are distributed smoothly over the 3D space, allowing a fit of the 3D
distribution to estimate backgrounds directly from the data.  A
potential signal would appear as a concentration of events in a small
region of the space (a 3D peak). We use a binned likelihood defined as
a function of parameters $m_{a_1}$, $m_{h_1}$ and effective signal
cross-section $\sigma \times \mathcal{B} (h_1 \to a_1 a_1)
\mathcal{B}^2(a_1 \to \mu\mu)$ to fit the simulated data using either
background-only or signal-plus-background templates.  We estimate the
sensitivity of this analysis and present it in terms of the 95\%
C.L.\ exclusion levels for signal cross-section using a Bayesian
technique.

Our estimations show that for an early data search ($\mathcal{L}
\simeq 100$~pb$^{-1}$), the backgrounds are negligibe.  For an
analysis with higher luminosity, one can restore the zero-background
situation by adding an isolation requirement to one or both of the
di-muon pairs in the event.  Isolation can be defined by either
setting an upper bound on the sum of the transverse momenta of all
tracks in a cone around the reconstructed direction of the di-muon
pair, excluding two muon tracks, or by rejecting pairs with additional
tracks above a certain threshold.  For the analysis with $\mathcal{L}
= 1$~fb$^{-1}$ of data, we required no charged tracks with momentum
$p_T > 1$~GeV/$c$ in the $\sqrt{(\Delta \eta)^2 + (\Delta \phi)^2} <
0.3$ cone around the direction of at least one of the two muon pairs.
This requirement is 96\% efficient for signal and reduces QCD multijet
background, dominated by events with muons originating from heavy
flavor jets, by a factor of 6--7.  For high-luminosity datasets,
isolation can be further tightened to increase background suppression
with only a moderate loss in signal efficiency.

\begin{table}[t]
\caption{Expected rate of background events per 100~pb$^{-1}$ of
  luminosity after selection. \label{bckgr_cuts_number_reco_level}}
\begin{center}
\renewcommand{\arraystretch}{1.4}
\begin{tabular}{c|c|c}
\hline\hline
Selection & 4 leptons & QCD multi-jet \\ 
\hline
$p_T (\mu_1) > 20$~GeV/$c$ and       &                           &                        \\
$p_T (\mu_i) > 5$~GeV/$c$; $i=2,3,4$ &               $4.8\pm0.2$ &             $267\pm23$ \\
\hline
$m_{12}$, $m_{34} < 4$~GeV/$c^2$     &           $0.024\pm0.012$ &              $90\pm13$ \\
$m_{1234} > 60$~GeV/$c^2$            &           $0.010\pm0.007$ &               $39\pm9$ \\
$|m_{12} - m_{34}| < 0.08$~GeV/$c^2$ &                           &                        \\
$+0.005 \times (m_{12} + m_{34})$    & $0.000^{+0.005}_{-0.000}$ & $0.00^{+1.95}_{-0.00}$ \\
\hline\hline
\end{tabular}
\end{center}
\end{table}

\begin{figure*}[htbp]
\includegraphics[width=0.32\linewidth]{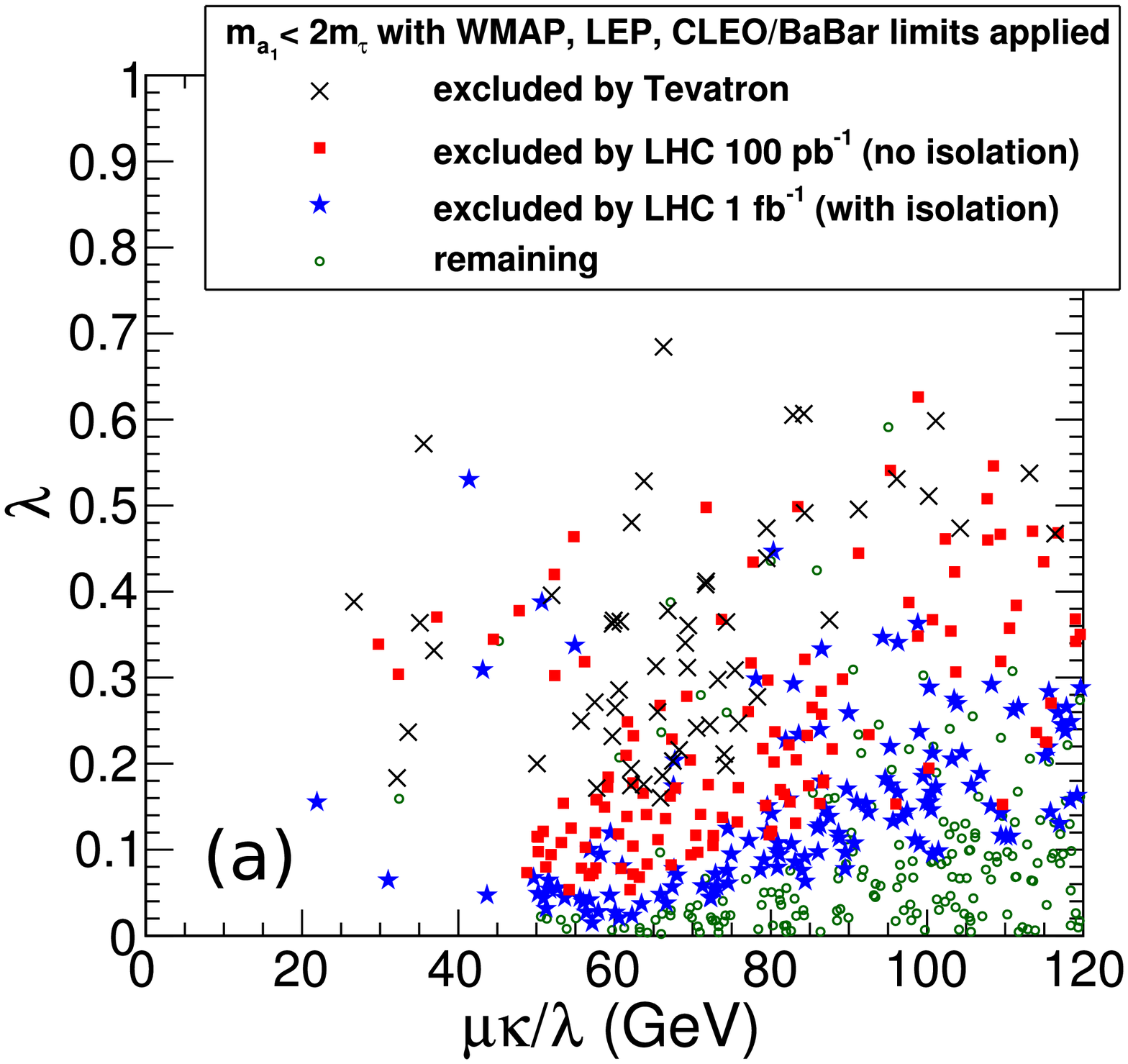}
\hfill
\includegraphics[width=0.32\linewidth]{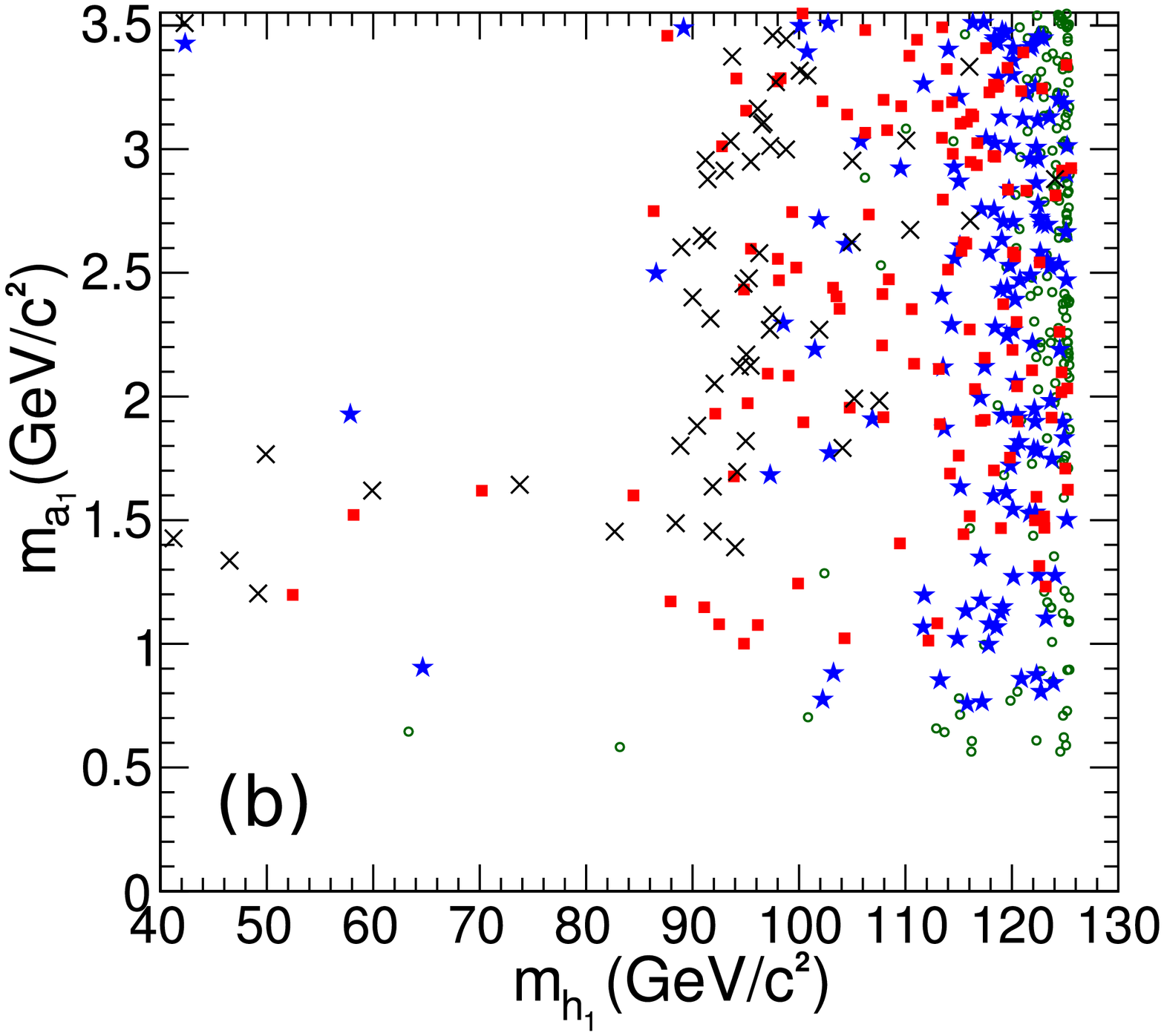}
\hfill
\includegraphics[width=0.32\linewidth]{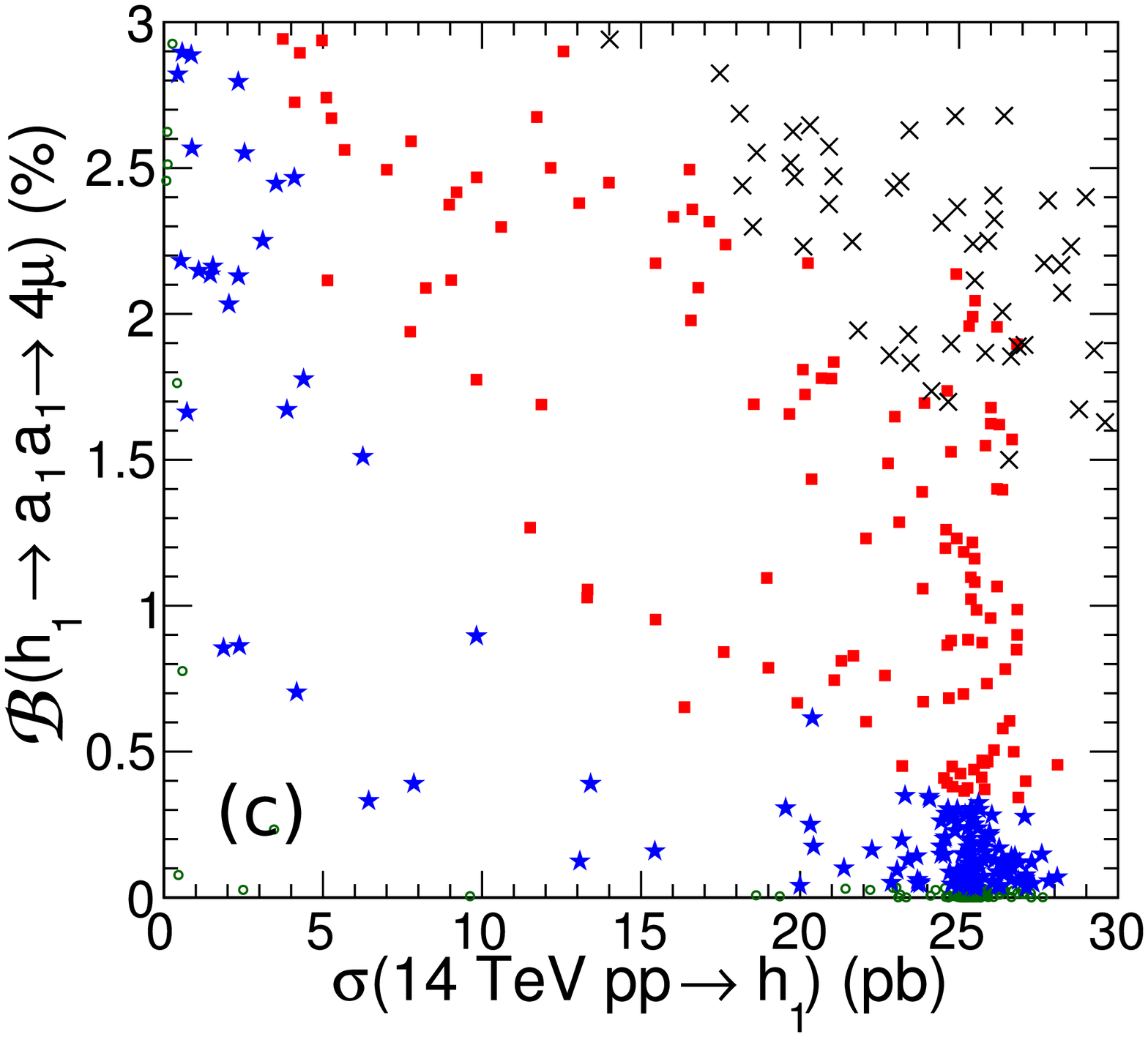}

\caption{Sampled models with $m_{a_1} < 2m_\tau$ and all experimental
  constraints applied, presented as a function of model parameters
  (a), Higgs masses (b), branching fraction and LHC cross-section (c).
  With only 100~pb$^{-1}$, the LHC's reach extends beyond that of the
  Tevatron. \label{fig:lhcexclusion}}
\end{figure*}

\subsection{LHC Reach for NMSSM $h_1 \to a_1 a_1$}

We have proposed an analysis that has a potential of discovering the
NMSSM with early LHC data in scenarios with low $m_{a_1}$.  We
estimate its sensitivity by calculating the 95\% C.L.\ upper limit on
the product $\sigma(pp \to h) \mathcal{B}(h_1 \to a_1 a_1)
\mathcal{B}^2(a_1 \to \mu\mu) \, \alpha$, where $\alpha$ is the
analysis acceptance, using a Bayesian technique.  Because of the low
background, an upper limit on the signal corresponds to approximately
three reconstructed events.  This limit is 0.0293~pb for $\mathcal{L}
= 100$~pb$^{-1}$, and scales linearly with luminosity assuming that
the number of observed background events is zero.  In nearly all
pseudoexperiments, this limit is independent of $m_{h_1}$ and
$m_{a_1}$ because the effective signal region that dominates signal
significance is essentially background-free and the probability to
observe an event is small.  Note that the corresponding projection for
$\mathcal{L} = 1$~fb$^{-1}$ includes the isolation cut, slightly
reducing signal efficiency and correspondingly loosening the limit.
The upper limit on $\sigma(pp \to h_1) \mathcal{B}(h_1 \to a_1 a_1)$ is
shown as a function of $m_{h_1}$ and $m_{a_1}$ in
Table~\ref{table_both_factorized}.

Figure~\ref{fig:lhcexclusion} presents the region of NMSSM parameter
space excluded by the Tevatron and the region that the LHC would
exclude with 100~pb$^{-1}$ (without isolation) and 1~fb$^{-1}$ (with
isolation), assuming no observed signal.  The regions are presented in
the $\lambda$, $\mu\kappa/\lambda$ plane (Fig.~11(a)), the $m_{a_1}$,
$m_{h_1}$ plane (Fig.~11(b)), and the plane of $h_1 \to a_1 a_1 \to
4\mu$ branching fraction versus LHC $pp \to h_1$ cross-section
(Fig.~11(c)).  High $h_1$-singlet scenarios (which have low production
cross-section) and low $h_1$-singlet scenarios (which have low $h_1
\to a_1 a_1$ branching fractions) are accessible to the Tevatron and
the LHC to different degrees, leading to a region in
Fig.~\ref{fig:lhcexclusion}(c) where high $h_1$-singlet scenarios are
excluded by the Tevatron while some low $h_1$-singlet models with the
same LHC cross-section times branching fraction are not.  The Tevatron
exclusion region has a sharp border only when viewed as a function of
the Tevatron cross-section.

\begin{table}[htbp]
\caption{95\% C.L.\ upper limit on $\sigma(pp \to h_1) \times
  \mathcal{B}(h_1 \to a_1 a_1 \to 4\mu)$ (fb) at the LHC with
  $\mathcal{L} = 100$~pb$^{-1}$ (no isolation) and $\mathcal{L} =
  1$~fb$^{-1}$ (with isolation).  The limit tightens at high $m_{h_1}$
  because of the increase in acceptance with
  $m_{h_1}$. \label{table_both_factorized}}

\begin{flushright}
\renewcommand{\arraystretch}{1.4}
\begin{tabular}{c | c | c | c | c}
\hline\hline
\multicolumn{5}{c}{$m_{a_1}$ (GeV/$c^2$)} \\
\hline
\mbox{\hspace{0.19 cm}}0.5\mbox{\hspace{0.19 cm}} & \mbox{\hspace{0.25 cm}}1.0\mbox{\hspace{0.25 cm}} & \mbox{\hspace{0.34 cm}}2.0\mbox{\hspace{0.34 cm}} & \mbox{\hspace{0.33 cm}}3.0\mbox{\hspace{0.33 cm}} & \mbox{\hspace{0.25 cm}}4.0\mbox{\hspace{0.25 cm}} \\
\end{tabular}

\begin{tabular}{c c | c | c | c | c}
\hline\hline \multicolumn{1}{c |}{$\mathcal{B}(a_1 \to \mu\mu)$ (\%)} & \mbox{\hspace{0.25 cm}}0\footnote{Recall that $\mathcal{B}(h_1 \to a_1 a_1)$ is obatined
using NMSSMTools and is not reliable for $m_{a_1} \lesssim 1$~GeV/$c^2$. Furthermore, this branching fraction is expected
to reach nearly 100\% for $2m_\mu < m_{a_1} < 3m_\tau$.}\mbox{\hspace{0.25 cm}} & \mbox{\hspace{0.25 cm}}9.8\mbox{\hspace{0.25 cm}} & \mbox{\hspace{0.25 cm}}15.2\mbox{\hspace{0.25 cm}} & \mbox{\hspace{0.25 cm}}16.2\mbox{\hspace{0.25 cm}} & \mbox{\hspace{0.25 cm}}0.7\mbox{\hspace{0.25 cm}} \\\hline\hline
& \multicolumn{5}{c}{for $\mathcal{L} = 100$~pb$^{-1}$ (no isolation)} \\
\hline
\multicolumn{1}{c |}{$m_{h_1} = 80$~GeV/$c^2$}  &  96.0 & 110.3 & 121.1 & 122.6 & 126.1 \\
\multicolumn{1}{c |}{$m_{h_1} = 100$~GeV/$c^2$} &  74.8 &  90.3 & 100.8 & 102.4 & 103.9 \\
\multicolumn{1}{c |}{$m_{h_1} = 120$~GeV/$c^2$} &  63.9 &  77.4 &  86.0 &  90.8 &  94.4 \\\hline\hline
& \multicolumn{5}{c}{for $\mathcal{L} = 1000$~pb$^{-1}$ (with isolation)} \\
\hline
\multicolumn{1}{c |}{$m_{h_1} = 80$~GeV/$c^2$}  &  10.0 &  11.5 &  12.6 &  12.8 &  13.1 \\
\multicolumn{1}{c |}{$m_{h_1} = 100$~GeV/$c^2$} &   7.8 &   9.4 &  10.5 &  10.7 &  10.8 \\
\multicolumn{1}{c |}{$m_{h_1} = 120$~GeV/$c^2$} &   6.7 &   8.1 &   9.0 &   9.5 &   9.8 \\\hline\hline
\end{tabular}
\end{flushright}
\end{table}

It is worth noting that quantitative background estimates performed in
our analysis may indicate that the LHC reach for NMSSM models with
$m_{a_1} > 2m_\tau$ in the $2\mu 2\tau$ channel are substantially
weaker than suggested in Ref.~\cite{2mu2tau-pheno}, which relied on
extrapolating QCD backgrounds to avoid high-statistics simulations.
Though the $4\mu$ and $2\mu 2\tau$ analyses apply different
selections, a rough extrapolation of the simulated QCD multijet
backgrounds to the $4\mu$ channel yields an estimate of backgrounds to
the $2\mu 2\tau$ channel that is three orders of magnitude larger than
what was used in Ref.~\cite{2mu2tau-pheno}, even without considering
the much larger misidentification rate of hadronically decaying taus.
The expected number of QCD multijet events in the $m_{1234} > 60$,
$m_{12}$ and $m_{34} < 4$~GeV/$c^2$ region of the $4\mu$ analysis,
which is $390 \pm 90$~events/fb$^{-1}$ (note that the numbers in
Table~\ref{bckgr_cuts_number_reco_level} are for 100~pb$^{-1}$), could
be reduced by a factor of 10--20 using tight isolation requirements.
Unlike our analysis, the study in Ref.~\cite{2mu2tau-pheno} applies
restrictions on the transverse momentum of the di-muon pair,
$p_T^{\mu\mu} > 40$~GeV/$c$, and on the missing transverse energy
$\met > 30$~GeV.  However, the $p_T^{\mu\mu}$ selection is similar to
$m_{1234} > 60$~GeV/$c^2$ and, given typical expected $\met$
resolution for multi-jet events, a $\met > 30$~GeV requirement cannot
be powerful enough to overcome several orders of magnitude in
estimated event count.  Scaling the $4\mu$ background estimate down by
a factor of 10 to account for muon isolation, we expect about $39 \pm
9$~events/fb$^{-1}$ from QCD multijets, as opposed to the
0.03~events/fb$^{-1}$ in Ref.~\cite{2mu2tau-pheno}.  Allowing for the
larger rate of tau misidentification compared to muons only increases
the discrepancy.  Another argument can be made using the $D\O$
measurement in the $2 \mu 2\tau$ channel~\cite{d0-low-ma}, which had
1--2 expected background events in the narrow ($\pm
0.3$--$1.0$~GeV/$c^2$) windows around the selected points in the
di-muon mass.  Using common scaling estimates of background
contributions from the Tevatron to the LHC, one would expect similar
backgrounds for LHC datasets of the order of 100~pb$^{-1}$.  For
datasets with integrated luminosities of the order of 500~pb$^{-1}$,
the corresponding QCD contamination would be 5--10 events per window.
These much larger background estimates severely affect acievable
exclusion limits and, considering trial factors, would make any
discovery in the $2\mu 2\tau$ channel with early data extremely
challenging.

\section{Conclusions}

We have studied the phenomenology of the NMSSM scenarios with the mass
of the lowest CP-odd Higgs boson, $a_1$, below the $2\tau$ threshold.
Our analysis of the impact of existing data on these models has shown
that the WMAP and LEP-II data provide the most constraining power,
while recent CLEO and BaBar measurements have essentially no impact on
the allowed parameter space, and the Tevatron data has only a weak
impact on the allowed parameter space.  As a result, a large fraction
of the parameter space is not excluded by any existing data.  We
conclude that a new analysis should be performed at the LHC to
definitively confirm or exclude these models.  We propose an analysis
suitable for the LHC using the $4\mu$ signature, which has very low
backgrounds and striking kinematical features allowing direct and
precise measurement of the masses of the $a_1$ and $h_1$ bosons.
Using the CMS experiment as a benchmark, we we estimate the sensitvity
of such an analysis and demonstrate that it has the potential to
either make a discovery or significantly diminish the allowed
parameter space of the NMSSM with low $m_{a_1}$ using only
100-1000~pb$^{-1}$ of early LHC data.
  
\section*{Acknowledgments}

We thank Ulrich Ellwanger, Cyril Hugonie, Alexander Pukhov, Jay Wacker
for useful discussions. One of the authors (AB) would like to thank
the GGI Institute (Florence), and another (JP) would like to thank
Fermilab, where important parts of the work on the paper were
performed.  This work would not be possible without the support of the
U.S. Department of Energy, the State of Texas, and the Universities
Research Association Visiting Scholars Program.

\def\Journal#1#2#3#4{{#1} {\bf #2}, #3 (#4)}
\def\NCA{Nuovo Cimento}
\def\NIM{Nucl. Instrum. Methods}
\def\NIMA{{Nucl. Instrum. Methods} A}
\def\NP{Nucl. Phys.} 
\def\NPB{{Nucl. Phys.} B}
\def\PLB{{Phys. Lett.}  B}
\def\PRL{Phys. Rev. Lett.}
\def\RPP{Rep. Prog. Phys.}
\def\PRD{{Phys. Rev.} D}
\def\PR{Phys. Rep.}
\def\PRP{Prog. Theor. Phys.}
\def\ZPC{{Z. Phys.} C}
\def\MPL{{Mod. Phys. Lett.} A}
\def\EPJC{{Eur. Phys. J.} C}
\def\CPC{Comput. Phys. Commun.}

\renewcommand{\baselinestretch}{1}

\end{document}